\documentclass[12pt]{article}
\usepackage{amsfonts, amsmath}

\evensidemargin\oddsidemargin
\newtheorem{theorem}{Theorem}
\newtheorem{lemma}{Lemma}

\newtheorem{definition}{Definition}
\newtheorem{corollary}[theorem]{Corollary}
\newtheorem{remark}{Remark}
\newenvironment{proof}[1][Proof]{\textbf{#1.} }{\ \hfill $\Box$ \vspace{0.5cm}}

\begin{document}

\vspace{2cm}

\begin{center}
{\LARGE Ultracoherence and Canonical Transformations}

\vspace{1cm}

{\large Joachim Kupsch}\footnote{%
e-mail: kupsch@physik.uni-kl.de}{\large \ and Subhashish Banerjee}\footnote{%
e-mail: subhashishb@rediffmail.com}

{\large Fachbereich Physik, TU Kaiserslautern\\[0pt]
D-67653 Kaiserslautern, Germany}
\end{center}

\vspace{0.5cm}

{\small The (in)finite dimensional symplectic group of homogeneous canonical
transformations is represented on the bosonic Fock space by the action of
the group on the ultracoherent vectors, which are generalizations of the
coherent states.}




\section{Introduction}

The linear canonical transformations are an important tool to study the
structure and the dynamics of quantum systems. An incomplete list of the
literature on this subject is \cite{AY:1982, Bargmann:1970, Berezin:1966,
Friedrichs:1953, Itzykson:1967, KMTP:1967, Ottesen:1995, Ruijsenaars:1978,
Shale:1962}. The aim of this paper is to give a self-contained presentation
of canonical transformations in quantum mechanics and in Fock space quantum
field theory using ultracoherent vectors. These vectors are generated by the
group of all linear canonical transformations acting on the
vacuum; they are Gaussian pure states and include the well known coherent
vectors and the squeezed vacua of quantum optics. The name \textit{%
ultracoherence} is taken from \cite{Slowikowski:1988}, where canonical
transformations are investigated with an algebra of normal ordered operators.

There is an extensive literature about the representations of the finite
dimension symplectic group. The special role of exponential vectors is
already emphasized in the publications of Bargmann \cite{Bargmann:1970} and
Itzykson \cite{Itzykson:1967} using the complex wave representation (or
Bargmann-Segal-Fock representation) of the Fock space. These authors use the
reproducing kernel property of the complex wave representation and construct
the kernel functions for the operators of the representation, see also \cite
{KMS:1975}. Our basic ansatz in Sect. \ref{ansatz} is motivated by formulas
in \cite{Bargmann:1970, Itzykson:1967, KMS:1975}. But the spirit of our
construction is closer to \cite{SSM:1988}. These authors investigate the
action of the symplectic group on Gaussian pure states in the
Schr\"{o}dinger representation of quantum mechanics. Transferred to the Fock
space language, the Gaussian pure states are the ultracoherent states for
the quantum mechanics of a finite number of degrees of freedom.

The representation of the infinite dimensional symplectic group on the
bosonic Fock space have been studied in \cite{Berezin:1966} using a symbolic
calculus related to the complex wave representation and in \cite
{Ruijsenaars:1978} with rigorous normal ordering expansions. The aim of our
paper is to construct the representation on a minimal set of vectors in the
Fock space, which is stable against Weyl transformations and homogeneous
canonical transformations. The linear span of these vectors, the
ultracoherent vectors, provide a natural domain for the (possibly unbounded)
generators of one-parameter subgroups of the infinite dimensional symplectic
group.

The plan of the paper is as follows. In Section \ref{Fock} we briefly
discuss Hilbert and bosonic Fock spaces. We define and discuss the
exponential vectors -- related to coherent states -- and the more general
ultracoherent vectors. In Section \ref{Weyl} we introduce the algebra of
Weyl operators, which defines the canonical structure with bounded
operators. In Section \ref{sympl} we investigate the symplectic group on
finite and infinite dimensional Hilbert spaces. In the case of infinite
dimensions the symplectic transformations can be implemented by unitary
operators on the Fock space only if an additional Hilbert-Schmidt condition
is satisfied \cite{Friedrichs:1953, Shale:1962}. Under this restriction we
construct in Section \ref{rep} a unitary ray representation of the
symplectic group on the bosonic Fock space by defining the action of this
group on exponential and ultracoherent vectors. The intertwining relations
of this representation with the algebra of Weyl operators are given in
Section \ref{Bogol}. In the concluding Section \ref{concl} we indicate
possible applications of our approach. Some detailed calculations for
ultracoherent vectors are given in the Appendix \ref{uv}.

\section{Fock space and ultracoherent vectors\label{Fock}}

\subsection{Hilbert spaces and Fock spaces\label{Hilbert}}

In this section we recapitulate some basic notations about Hilbert spaces
and Fock spaces of symmetric tensors. Let $\mathcal{H}$ be a complex
separable Hilbert space with inner product $\left( f\mid g\right) $ and with
an antiunitary involution $f\rightarrow f^{\ast },\,f^{\ast \ast }\equiv
(f^{\ast })^{\ast }=f$. Then the mapping
\begin{equation}
f,\,g\rightarrow \left\langle f\mid g\right\rangle :=\left( f^{\ast }\mid
g\right) \in \mathbb{C}  \label{h1}
\end{equation}
is a symmetric bilinear form $\left\langle f\mid g\right\rangle
=\left\langle g\mid f\right\rangle $. The underlying real Hilbert space of $%
\mathcal{H}$ is denoted as $\mathcal{H}_{\mathbb{R}}$. This space has the
inner product $\left( f\mid g\right) _{\mathbb{R}}=\mathrm{Re}\,\left( f\mid
g\right) \newline
=\frac{1}{2}\left( \left( f\mid g\right) +\left( f^{\ast }\mid g^{\ast
}\right) \right) $. As a point set the spaces $\mathcal{H}$ and $\mathcal{H}%
_{\mathbb{R}}$ coincide. For some calculations it is advantageous to
identify $\mathcal{H}_{\mathbb{R}}$ with diagonal subspace $\mathcal{H}%
_{diag}$ of $\mathcal{H}\times \mathcal{H}^{\ast }.$ This space is defined
as the set of all elements $\left(
\begin{array}{c}
f \\
g^{\ast }
\end{array}
\right) \in \mathcal{H}\times \mathcal{H}^{\ast }$ which satisfy $g=f$.

We use the following notations for linear operators: the space of all
bounded operators $A$ with operator norm $\left\| A\right\| $ is $\mathcal{L}%
(\mathcal{H})$, the space of all Hilbert-Schmidt operators $A$ with norm $%
\left\| A\right\| _{HS}=\sqrt{\mathrm{tr}_{\mathcal{H}}A^{+}A}$ is $\mathcal{%
L}_{2}(\mathcal{H})$, the space of all trace class or nuclear operators $A$
with norm $\left\| A\right\| _{1}=\mathrm{tr}_{\mathcal{H}}\sqrt{A^{+}A}$ is
$\mathcal{L}_{1}(\mathcal{H})$. For operators $A\in \mathcal{L}(\mathcal{H})$
the adjoint operator is denoted by $A^{+}$. The complex conjugate operator $%
\bar{A}$ and the transposed operator $A^{T}$ are defined by the identities
\begin{equation}
\bar{A}f=\left( Af^{\ast }\right) ^{\ast },\;A^{T}f=\left( A^{+}f^{\ast
}\right) ^{\ast }  \label{h2}
\end{equation}
for all $f\in \mathcal{H}$. The usual relations $A^{+}=\left( \bar{A}\right)
^{T}=\overline{\left( A^{T}\right) }$ are valid. An operator $A$ with the
property $A=A^{T}$ is called a \textit{transposition-symmetric operator}. It
satisfies the symmetry relation $\left\langle f\mid Ag\right\rangle
=\left\langle Af\mid g\right\rangle $ for all $f,g\in \mathcal{H}$.

Let $\mathcal{H}^{\otimes n},\,n\in \mathbb{N},$ be the algebraic n-th
tensor power of the Hilbert space $\mathcal{H}.$ The norm of $\mathcal{H}%
^{\otimes n}$ is fixed with the normalization $\left\| f_{1}\otimes
f_{2}\otimes ...\otimes f_{n}\right\| _{n}=\sqrt{n!}\prod_{j=1}^{n}\left\|
f_{j}\right\| $ for the product of $n$ vectors $f_{j}\in \mathcal{H}$. The
completion of $\mathcal{H}^{\otimes n},\,n\geq 2,$ with this norm is the
Hilbert space $\widehat{\mathcal{H}^{\otimes n}}$. The projection operator $%
P_{n}:$ $\widehat{\mathcal{H}^{\otimes n}}\rightarrow \widehat{\mathcal{H}%
^{\otimes n}}$ onto symmetric tensors of degree $n$ is uniquely given by the
prescription $P_{n}\left( f_{1}\otimes f_{2}\otimes ...\otimes f_{n}\right) =%
\newline
(n!)^{-1}\sum_{\sigma }f_{\sigma (1)}\otimes f_{\sigma (2)}\otimes
...\otimes f_{\sigma (n)}$ where $\sigma $ runs over all permutations of the
numbers $\left\{ 1,...,n\right\} $. The algebraic space of symmetric tensors
is $\mathcal{H}^{\vee n}=P_{n}\mathcal{H}^{\otimes n}$ and the completion is
the Hilbert space $\widehat{\mathcal{H}^{\vee n}}=P_{n}\widehat{\mathcal{H}%
^{\otimes n}}$. With $\widehat{\mathcal{H}^{\otimes 0}}=\mathbb{C}$ and $%
\widehat{\mathcal{H}^{\otimes 1}}=\mathcal{H}^{\otimes 1}=\mathcal{H}$ the
Fock space of all tensors is the direct sum $\mathcal{T}(\mathcal{H}\mathbb{)%
}=\oplus _{n=0}^{\infty }\widehat{\mathcal{H}^{\otimes n}}$ where the norm
is defined by $\left\| F\right\| ^{2}=\sum_{n=0}^{\infty }\left\|
F_{n}\right\| _{n}^{2}$ if $F=\sum_{n=0}^{\infty }F_{n},\,F_{n}\in \widehat{%
\mathcal{H}^{\otimes n}}$. With $\widehat{\mathcal{H}^{\vee 0}}=\mathbb{C}$
and $\widehat{\mathcal{H}^{\vee 1}}=\mathcal{H}^{\vee 1}=\mathcal{H}$ the
Hilbert sum $\mathcal{S}(\mathcal{H})=\oplus _{n=0}^{\infty }\widehat{%
\mathcal{H}^{\vee n}}$ defines the Fock space of symmetric tensors as
subspace of $\mathcal{T}(\mathcal{H}\mathbb{)}$. The symmetric tensor
product of the tensors $F\in \widehat{\mathcal{H}^{\vee m}}$ and $G\in
\widehat{\mathcal{H}^{\vee n}}$ is defined as $F\vee G=P_{m+n}(F\otimes G)$.
Then the usual normalization $\left\| f_{1}\vee ...\vee f_{n}\right\|
_{n}^{2}=\mathrm{per}\left( (f_{k}\mid f_{l})\right) $ with the permanent
follows for the norm of the product of $n$ vectors $f_{j}\in \mathcal{H}%
,\,j=1,...,n$.

Since $\left\| F\otimes G\right\| _{m+n}\leq \sqrt{\frac{(m+n)!}{m!n!}}%
\left\| F\right\| _{m}\left\| G\right\| _{n}$ with the norm introduced
above, we also have
\begin{equation}
\left\| F\vee G\right\| _{m+n}\leq \sqrt{\frac{(m+n)!}{m!n!}}\left\|
F\right\| _{m}\left\| G\right\| _{n}.  \label{h3a}
\end{equation}
By linear extension the symmetric tensor product is extended to the
algebraic sum $\mathcal{S}_{fin}(\mathcal{H})=\oplus _{n}\widehat{\mathcal{H}%
^{\vee n}}$ (linear subset of all $F\in \mathcal{S}(\mathcal{H})$ which have
components in a finite number of subspaces $\widehat{\mathcal{H}^{\vee n}}$
only). The space $\mathcal{S}_{fin}(\mathcal{H})$ is an algebra with respect
to the symmetric tensor product; the unit is the normalized basis vector $%
\mathbf{1}_{vac}$ (vacuum) of the space $\widehat{\mathcal{H}^{\vee 0}}=%
\mathbb{C}$. If we restrict the Hilbert spaces $\widehat{\mathcal{H}^{\vee n}%
}$ to the algebraic tensor spaces $\mathcal{H}^{\vee n}$ we obtain the
algebra $\mathcal{S}_{_{alg}}^{0}(\mathcal{H})=\oplus _{n}\mathcal{H}^{\vee
n}$, which is strictly smaller than $\mathcal{S}_{fin}(\mathcal{H})$ if $%
\dim \mathcal{H=\infty }$, but still dense in $\mathcal{S}(\mathcal{H})$.

The inner product of two elements $F,G$ of $\mathcal{S}(\mathcal{H})$ is
written as $\left( F\mid G\right) $. The antiunitary involution $%
f\rightarrow f^{\ast }$ on $\mathcal{H}$ can be uniquely extended to an
antiunitary involution $F\rightarrow F^{\ast }$on $\mathcal{S}(\mathcal{H})$
with the rule $\left( F\vee G\right) ^{\ast }=G^{\ast }\vee F^{\ast
}=F^{\ast }\vee G^{\ast }$. The mapping $F,\,G\in \mathcal{S}(\mathcal{H}%
)\rightarrow \left\langle F\mid G\right\rangle :=\left( F^{\ast }\mid
G\right) \in \mathbb{C}$ is again a $\mathbb{C}-$bilinear symmetric form.

The normalizations of the symmetric tensor product used in this paper agree
with those of \cite{Nielsen:1991}. The algebra $\Gamma _{0}\mathcal{H}$ of
\cite{Nielsen:1991} coincides with the algebra $\mathcal{S}_{_{alg}}^{0}(%
\mathcal{H})$ defined above.

\subsection{Exponential vectors\label{exp}}

For all vectors $f\in \mathcal{H}$ the exponential series $\exp f=1_{vac}+f+%
\frac{1}{2!}f\vee f+...$ is absolutely summable within $\mathcal{S}(\mathcal{%
H})$ and it satisfies the usual factorization property $\exp f\vee \exp
g=\exp (f+g)$, see \cite{Nielsen:1991} and also Appendix \ref{norm} of this
paper. The mapping $f\rightarrow \exp f$ is an entire analytic function%
\footnote{%
In this article analyticity means the existence of norm convergent power
series expansions as used e.g. in \cite{Hille/Phillips:1957}.}. The inner
product of two exponential vectors is
\begin{equation}
\left( \exp f\mid \exp g\right) =\exp \left( f\mid g\right) .  \label{h4}
\end{equation}
Coherent states are the normalized exponential vectors $\exp \left( f-\frac{1%
}{2}\left\| f\right\| ^{2}\right) \in \mathcal{S}(\mathcal{H})$. The linear
span of all exponential vectors $\left\{ \exp f\mid f\in \mathcal{H}\right\}
$ will be denoted by $\mathcal{S}_{coh}(\mathcal{H})$. The involution of an
exponential vector is $\left( \exp f\right) ^{\ast }=\exp f^{\ast },\,f\in
\mathcal{H}$. Due to the factorization property $\exp f\vee \exp g=\exp
(f+g) $ the set $\mathcal{S}_{coh}(\mathcal{H})$ is an algebra.

\begin{lemma}
\label{dense}The set $\left\{ \exp f\mid f\in \mathcal{H}\right\} $ of all
exponential vectors is linearly independent and the linear span $\mathcal{S}%
_{coh}(\mathcal{H})$ of these vectors is dense in $\mathcal{S}(\mathcal{H})$.
\end{lemma}

\begin{proof}
A proof is given in \cite{Guichardet:1972} \S\ 2.1 and in \cite
{Parthasarathy:1992} Proposition 19.4.
\end{proof}

A (bounded) operator on $\mathcal{S}(\mathcal{H})$ it is therefore uniquely
determined, if this operator is known on all exponential vectors. Actually
it is sufficient to define the operator on a set $\left\{ \exp f\mid f\in
\mathcal{D}\right\} $ where $\mathcal{D}$ is dense in $\mathcal{H}$, see
Corollary 19.5 of \cite{Parthasarathy:1992}. As example a bounded operator $%
B\in \mathcal{L}(\mathcal{H})$ can be lifted to an operator $\Gamma (B)$ on $%
\mathcal{S}(\mathcal{H})$ using the prescription
\begin{equation}
\Gamma (B)\exp f:=\exp Bf,~f\in \mathcal{H}.  \label{h7}
\end{equation}
The operator $\Gamma (B)$ is continuous, if $B$ is a contraction, i. e. $%
\left\| B\right\| \leq 1$; and it is isometric/unitary, if $B$ is
isometric/unitary. The statement about isometry is an immediate consequence
of

\begin{lemma}
\label{isom}Let $T_{0}$ be a linear operator $T_{0}:\mathcal{S}_{coh}(%
\mathcal{H})\rightarrow \mathcal{S}(\mathcal{H})$ which satisfies
\begin{equation}
\left( T_{0}\exp f\mid T_{0}\exp g\right) =\left( \exp f\mid \exp g\right)
=\exp \left( f\mid g\right)  \label{h5}
\end{equation}
for all $f,g\in \mathcal{H}$, then $T_{0}$ can be uniquely extended to a
linear isometric mapping $T$ on $\mathcal{S}(\mathcal{H})$.
\end{lemma}

\begin{proof}
The proof follows from Lemma \ref{dense} and from the Proposition 7.2 of
\cite{Parthasarathy:1992}.
\end{proof}

We have already stated that the dense linear subsets $\mathcal{S}_{fin}(%
\mathcal{H})$ and $\mathcal{S}_{coh}(\mathcal{H})$ are algebras with respect
to the symmetric tensor product. Take $F\in \mathcal{S}_{fin}(\mathcal{H})$
and $G\in \mathcal{S}_{coh}(\mathcal{H})$ then the product $F\vee G=G\vee F$
is defined, and the linear span of these products $\mathcal{S}_{_{alg}}(%
\mathcal{H})=\newline
span\,\left\{ F\vee \exp g\mid F\in \mathcal{S}_{fin}(\mathcal{H}),\,g\in
\mathcal{H}\right\} $ is again an algebra. The proof of the corresponding
statement has been given for the algebra $\mathcal{S}_{_{alg}}^{0}(\mathcal{H%
})\subset \mathcal{S}_{_{alg}}(\mathcal{H})$ generated by $\mathcal{S}%
_{_{fin}}^{0}(\mathcal{H})$ and $\mathcal{S}_{coh}(\mathcal{H})$ in \cite
{Nielsen:1991} Theorem 3.4. The essential estimate for this proof is (\ref
{h3a}), which applies to $\mathcal{H}^{\vee n}$ and to $\widehat{\mathcal{H}%
^{\vee n}}$. The proof of \cite{Nielsen:1991} can therefore easily be
extended to the algebra $\mathcal{S}_{_{alg}}(\mathcal{H})$.

For all $f,\,g\in \mathcal{H}$ the product $f\vee \exp g$ is an element of $%
\mathcal{S}(\mathcal{H})$. Given a vector $f\in \mathcal{H}$ the creation
operator $a^{+}(f)$ and the corresponding annihilation operator $a(f)$ are
uniquely determined by
\begin{equation}
\begin{array}{c}
a^{+}(f)\exp g=f\vee \exp g \\
a(f)\exp g=\left\langle f\mid g\right\rangle \exp g.
\end{array}
\label{h8}
\end{equation}
These operators are related by $\left( a^{+}(f)\right) ^{+}=a(f^{\ast })$,
and they satisfy the canonical commutation relations $\left[ a(f),a^{+}(g)%
\right] =\left\langle f\mid g\right\rangle \,I,~\left[ a^{+}(f),a^{+}(g)%
\right] =\left[ a(f),a(g)\right] =0$ for all $f,\,g\in \mathcal{H}$. These
relations are equivalent to
\begin{equation}
\left[ a^{+}(f)-a(f^{\ast }),a^{+}(g)-a(g^{\ast })\right] =-2i\omega (f,g)\,I
\label{h12}
\end{equation}
with
\begin{equation}
\omega (f,g):=\frac{1}{2i}\left( \left\langle f^{\ast }\mid g\right\rangle
-\left\langle f\mid g^{\ast }\right\rangle \right) =\mathrm{Im}\left( f\mid
g\right) \in \mathbb{R},  \label{h13}
\end{equation}
which is an $\mathbb{R}$-bilinear continuous skew symmetric form on the
Hilbert space $\mathcal{H}$ or, more precisely, on the underlying real space
$\mathcal{H}_{\mathbb{R}}$. The creation and annihilation operators are
unbounded operators. A domain of definition, on which these operators and
their commutators are meaningful is the algebra $\mathcal{S}_{_{alg}}(%
\mathcal{H})$.

\subsection{Ultracoherent vectors}

The space $\mathcal{L}_{2}(\mathcal{H})$ of Hilbert-Schmidt operators on $%
\mathcal{H}$ is a Hilbert space with the Hilbert-Schmidt norm $\left\|
A\right\| _{HS}=\sqrt{\mathrm{tr}A^{+}A}$. The restriction to
transposition-symmetric Hilbert-Schmidt operators $\left\{ A\in \mathcal{L}%
_{2}(\mathcal{H})\mid A=A^{T}\right\} $ is a closed subspace of $\mathcal{L}%
_{2}(\mathcal{H})$. This space is called $\mathcal{L}_{2sym}(\mathcal{H})$.
There exists a linear isomorphism between $\mathcal{L}_{2sym}(\mathcal{H})$
and the Hilbert space $\widehat{\mathcal{H}^{\vee 2}}$ of tensors of second
degree:

\begin{lemma}
\label{iso}Let $A$ be an operator in $\mathcal{L}_{2sym}(\mathcal{H})$, then
there exists a unique tensor of second degree, in the sequel denoted by $%
\Omega (A)$, such that
\begin{equation}
\left\langle \Omega (A)\mid f\vee g\right\rangle =\left\langle f\mid
Ag\right\rangle =\left\langle g\mid Af\right\rangle  \label{h14}
\end{equation}
for all $f,\,g\in \mathcal{H}$. The mapping $A\in \mathcal{L}_{2sym}(%
\mathcal{H})\rightarrow \Omega (A)\in \widehat{\mathcal{H}^{\vee 2}}$ is an
invertible continuous linear transformation between the spaces $\mathcal{L}%
_{2sym}(\mathcal{H})$ and $\widehat{\mathcal{H}^{\vee 2}}$. The respective
norms are related by $\left\| \Omega (A)\right\| _{2}^{2}=\frac{1}{2}\left\|
A\right\| _{HS}^{2}$.
\end{lemma}

\begin{proof}
As $\left\langle \Omega (A)\mid f\vee g\right\rangle =0$ for all $f,\,g\in
\mathcal{H}$ implies $\Omega (A)=0$, the tensor $\Omega (A)$ is uniquely
determined by (\ref{h14}). For the construction of $\Omega (A)$ we choose a
real orthonormal basis $e_{\mu }=e_{\mu }^{\ast }$ of the Hilbert space $%
\mathcal{H}$. Then any transposition-symmetric Hilbert-Schmidt operator $A$
has matrix elements $A_{\mu \nu }=\left( e_{\mu }\mid A\,e_{\nu }\right)
=\left\langle e_{\mu }\mid A\,e_{\nu }\right\rangle =A_{\nu \mu }$ which are
square summable $\sum_{\mu \nu }\left| A_{\mu \nu }\right| ^{2}=\left\|
A\right\| _{HS}^{2}<\infty $. The tensor $\Omega (A):=\frac{1}{2}\sum_{\mu
\nu }A_{\mu \nu }\,e_{\mu }\vee e_{\nu }\in \widehat{\mathcal{H}^{\vee 2}}$
then satisfies the identity (\ref{h14}). The tensor norm of $\Omega (A)$ is
calculated as $\frac{1}{2}\sum_{\mu \nu }\left| A_{\mu \nu }\right| ^{2}$
and the norm identity follows.

On the other hand, given a tensor $F\in \widehat{\mathcal{H}^{\vee 2}}$ we
have $F=\frac{1}{2}\sum_{\mu \nu }F_{\mu \nu }\,e_{\mu }\vee e_{\nu }$ with
coefficients $F_{\mu \nu }=F_{\upsilon \mu }\in \mathbb{C}$, which are
square summable. Then the operator $f\in \mathcal{H}\rightarrow Af=\sum_{\mu
}F_{\mu \nu }\left\langle e_{\mu }\mid f\right\rangle \,e_{\upsilon }\in
\mathcal{H}$ is obviously a transposition-symmetric Hilbert-Schmidt operator
with $\Omega (A)=F$.
\end{proof}

\begin{definition}
\label{siegel}The set of all Hilbert--Schmidt operators $A\in \mathcal{L}%
_{2sym}(\mathcal{H})$ with an operator norm strictly less than one, $\left\|
A\right\| <1$, is called the Siegel (unit) disc. It is denoted by $\mathbf{D}%
_{1}$.
\end{definition}

For Hilbert spaces with finite dimensions this disc has been introduced by
Siegel \cite{Siegel:1943}. The set $\mathbf{D}_{1}$ is open and convex, and
it is stable against transformations $A\rightarrow UAU^{T}$ with a unitary
operator $U$. The last statement follows from $\mathrm{tr}_{\mathcal{H}}\bar{%
U}A^{+}U^{+}UAU^{T}=\mathrm{tr}_{\mathcal{H}}A^{+}A$ and $\left\|
UAU^{T}\right\| \leq \left\| A\right\| $.

In \cite{KMTP:1967} it has been derived that the exponential series
\begin{equation}
\exp \Omega (A)=1_{vac}+\Omega (A)+\frac{1}{2!}\Omega (A)\vee \Omega (A)+...
\label{h16}
\end{equation}
converges within the Fock space $\mathcal{S}(\mathcal{H})$ if $A\in \mathbf{D%
}_{1}$. The convergence is uniform for each subset $\mathbf{D}_{1}^{c,\delta
}=\left\{ A\in \mathbf{D}_{1}\mid \left\| A\right\| _{HS}\leq c<\infty
,\,\left\| A\right\| \leq \delta <1\right\} $. The mapping $A\in \mathbf{D}%
_{1}\rightarrow \exp \Omega (A)\in \mathcal{S}(\mathcal{H})$ is analytic.
The inner product of two of these exponentials can be calculated as
\begin{equation}
\left( \exp \Omega (A)\mid \exp \Omega (B)\right) =\left( \det {}_{\mathcal{H%
}}\left( I-A^{+}B\right) \right) ^{-\frac{1}{2}}=\left( \det {}_{\mathcal{H}%
}\left( I-BA^{+}\right) \right) ^{-\frac{1}{2}}.  \label{h17}
\end{equation}
The proof of this identity follows from Theorem 2 of \cite{KMTP:1967}, but
see also Appendix \ref{series}. The identity (\ref{h17}) implies that $\exp
\Omega (A)\in \mathcal{S}(\mathcal{H})$ if and only if $A\in \mathbf{D}_{1}$.

In Appendix \ref{norm} we prove that the symmetric tensor product of the
tensors $\exp \Omega (A),\newline
\,A\in \mathbf{D}_{1}$, and $\exp f,\,f\in \mathcal{H}$, is defined within
the Fock space $\mathcal{S}(\mathcal{H})$. For any operator $A\in \mathbf{D}%
_{1}$ and for any $f\in \mathcal{H}$ we now define the \textit{ultracoherent
vector}
\begin{equation}
\Phi (A,f):=\exp \Omega (A)\vee \exp f=\exp f\vee \exp \Omega (A)\in
\mathcal{S}(\mathcal{H}).  \label{h18}
\end{equation}
The result (\ref{h17}) can be extended to the inner product of two
ultracoherent vectors
\begin{equation}
\begin{array}{l}
\left( \Phi (A,f)\mid \Phi (B,g)\right) \\
=\left( \det {}_{\mathcal{H}}\left( I-A^{+}B\right) \right) ^{-\frac{1}{2}%
}\exp \left( \frac{1}{2}\left\langle f^{\ast }\mid Cf^{\ast }\right\rangle
+\left\langle f^{\ast }\mid (I-BA^{+})^{-1}\;g\right\rangle +\frac{1}{2}%
\left\langle g\mid D\;g\right\rangle \right)
\end{array}
\label{h19}
\end{equation}
with
\begin{equation}
\begin{array}{c}
C=B(I-A^{+}B)^{-1}=(I-BA^{+})^{-1}B \\
D=A^{+}(I-BA^{+})^{-1}=(I-A^{+}B)^{-1}A^{+}.
\end{array}
\label{h20}
\end{equation}
Proofs of this identity are given in the Appendices \ref{identities} and \ref
{series}.

\begin{remark}
\label{BF}For $F\in \mathcal{S}(\mathcal{H})$ the function
\begin{equation}
\Phi _{F}(z^{\ast })=(\exp z\mid F)=\left\langle \exp z^{\ast }\mid
F\right\rangle  \label{bf1}
\end{equation}
is entire antianalytic in the variable $z\in \mathcal{H}$, and the tensor $F$
is uniquely determined by this function. We denote the linear space of all
functions $\left\{ \Phi _{F}(z^{\ast })\mid F\in \mathcal{S}(\mathcal{H}%
)\right\} $ by $\mathcal{B}$. Then $\mathcal{B}$ can be equipped with the
Hilbert space topology induced by the topology of $\mathcal{S}(\mathcal{H})$%
, i.e., the inner product $\left( \Phi _{F}\parallel \Phi _{G}\right) $ of
two functions $\Phi _{F}$ and $\Phi _{G}$ is defined as $\left( \Phi
_{F}\parallel \Phi _{G}\right) :=\left( F\mid G\right) $. With this
structure the space $\mathcal{B}$ becomes a Hilbert space with the
reproducing kernel $\exp \left\langle z^{\ast }\mid w\right\rangle $. This
representation of the bosonic Fock space is called the complex wave
representation or Bargmann-Fock representation, see \cite{Bargmann:1961,
Segal:1962} and the more recent publications \cite{BSZ:1992, Hall:2000,
Nielsen:1991}. The exponential vectors and the ultracoherent vectors have a
simple representation in this space: $\exp f\in \mathcal{S}(\mathcal{H})$
corresponds to the exponential function $(\exp z\mid \exp f)=\exp
\left\langle z^{\ast }\mid f\right\rangle $ and the ultracoherent vectors
are given by (\ref{id3}). \newline
If $F\in \mathcal{S}_{coh}(\mathcal{H})$ then $\Phi _{F}(z^{\ast })$ is a
tame function, i.e., it depends only on a finite number of variables $%
\left\langle z^{\ast }\mid f_{j}\right\rangle ,\,f_{j}\in \mathcal{H}%
,\,j=1,...,N$. Let $\upsilon (dz,dz^{\ast })$ be the canonical Gaussian
promeasure on the Hilbert space $\mathcal{H}_{\mathbb{R}}$, then tame
functions can be integrated, and the identity
\begin{equation}
\int_{\mathcal{H}_{\mathbb{R}}}\overline{\Phi _{F}(z^{\ast })}\Phi
_{G}(z^{\ast })\upsilon (dz,dz^{\ast })=\left( F\mid G\right)  \label{bf2}
\end{equation}
holds for all $F,G\in \mathcal{S}_{coh}(\mathcal{H})$. If $A\in \mathcal{D}%
_{1}$ is a finite rank operator, then (\ref{id3}) is a tame function. The
integral (\ref{bf2}) can be used to calculate the inner product (\ref{h19})
-- first for finite rank operators $A$ and $B$ and then by a continuity
argument for general $A,B\in \mathcal{D}_{1}$. A calculation of the integral
for finite dimensional Hilbert spaces can be found in Appendix II of \cite
{Itzykson:1967}. The proof of (\ref{h19}), which we present in Appendix \ref
{uv}, does not use this technique.
\end{remark}

\section{Weyl operators and canonical transformations\label{Weyl}}

\subsection{Weyl operators}

In this Section we recapitulate some properties of Weyl operators needed for
the subsequent investigations. The Weyl operators $W(h),\,h\in \mathcal{H}$,
are defined on the linear span of the exponential vectors by
\begin{equation}
W(h)\exp f=\mathrm{e}^{-\left( h\mid f\right) -\frac{1}{2}\left\| h\right\|
^{2}}\exp (f+h)=\mathrm{e}^{-\left\langle h^{\ast }\mid f\right\rangle -%
\frac{1}{2}\left\langle h^{\ast }\mid h\right\rangle }\exp (f+h).  \label{w1}
\end{equation}
It is straightforward to derive the identity
\begin{equation*}
\left( W(h)\exp f\mid W(h)\exp g\right) =\mathrm{e}^{\left( f\mid g\right)
}=\left( \exp f\mid \exp g\right)
\end{equation*}
for $f,g,h\in \mathcal{H}$. Then Lemma \ref{isom} implies that $W(h)$ is
isometric. On the other hand we have $W(h)W(-h)=id$ and $W(h)$ is invertible
with $W^{-1}(h)=W(-h)$. The Weyl operators can therefore be extended to
unitary operators on the Fock space $\mathcal{S}(\mathcal{H})$. Calculating $%
W(f)W(g)\exp h$ and $W(f+g)\exp h$ we obtain the \textit{Weyl relations}
\begin{equation}
W(f)W(g)=\mathrm{e}^{-i\omega (f,g)}W(f+g)  \label{w3}
\end{equation}
with the skew symmetric form (\ref{h13}) on the underlying real space $%
\mathcal{H}_{\mathbb{R}}$. The identity (\ref{w3}) defines the canonical
structure on the Fock space $\mathcal{S}(\mathcal{H})$. The Weyl relations
are equivalent to the canonical commutation relations (\ref{h12}). The
advantage of the Weyl relations is that they are formulated with bounded
operators.

The action of the Weyl operator on the ultracoherent vector is calculated in
Appendix \ref{identities} as
\begin{equation}
W(h)\Phi (A,f)=\mathrm{e}^{-\frac{1}{2}\left\| h\right\| ^{2}+\frac{1}{2}%
\left\langle h^{\ast }\mid Ah^{\ast }-2f\right\rangle }\Phi (A,f+h-Ah^{\ast
}).  \label{w4}
\end{equation}

As well known, the Weyl operators have a simple representation in terms of
the creation and annihilation operators. Differentiating $W(\lambda h)\exp f$
with respect to $\lambda \in \mathbb{R}$ and comparing the result with (\ref
{h8}) we obtain the usual representation of the Weyl operator $W(h)=\exp
\left( a^{+}(h)-a(h^{\ast })\right) $. Using (\ref{w1}) together with (\ref
{h8}) we get the relations
\begin{equation}
\begin{array}{l}
W(h)a^{+}(f)W^{+}(h)=a^{+}(f)-\left( h\mid f\right) =a^{+}(f)-\left\langle
f\mid h^{\ast }\right\rangle , \\
W(h)a(f)W^{+}(h)=a(f)-\left( f^{\ast }\mid h\right) =a(f)-\left\langle f\mid
h\right\rangle .
\end{array}
\label{w5}
\end{equation}

\subsection{Canonical transformations\label{canonical}}

Canonical transformations are unitary operators $S$ on $\mathcal{S}(\mathcal{%
H})$ which preserve the canonical commutation relations (\ref{h12}). To
avoid any discussion about the domain of the operators $Sa^{+}(f)S^{+}$ and $%
Sa(f^{\ast })S^{+}$ we demand the invariance of the Weyl relations (\ref{w3}%
)
\begin{equation}
SW(f)S^{+}SW(g)S^{+}=SW(f)W(g)S^{+}=\mathrm{e}^{-i\mathrm{Im}\left( f\mid
g\right) }SW(f+g)S^{+}.  \label{w6}
\end{equation}

There are two types of linear canonical transformations:

\begin{enumerate}
\item  The inhomogeneous Transformations generate a c-number shift for the
creation and annihilation operators. These transformations are given by the
unitary Weyl operators, as can be seen from the relations (\ref{w5}). The
invariance of the Weyl relations
\begin{equation*}
W(h)W(f)W(g)W(-h)=\mathrm{e}^{-i\mathrm{Im}\left( f\mid g\right)
}W(h)W(f+g)W(-h)
\end{equation*}
with $S=W(h),\,h\in \mathcal{H}$, as canonical transformation easily follow
from (\ref{w3}).

\item  The homogeneous canonical transformations generate linear
transformations between the creation and annihilation operators
\begin{equation}
\begin{array}{l}
Sa^{+}(f)S^{+}=a^{+}(Uf)-a(\bar{V}f), \\
Sa(f)S^{+}=-a^{+}(Vf)+a(\bar{U}f).
\end{array}
\label{w8}
\end{equation}
Here $U$ and $V$ are bounded linear transformations on $\mathcal{H}$. The
relations (\ref{w8}) imply $S\left( a^{+}(f)-a(f^{\ast })\right)
S^{+}=a^{+}(Uf+Vf^{\ast })-a\left( \bar{V}f+\bar{U}f^{\ast }\right) $. The
canonical transformations of this type are usually called \textit{Bogoliubov
transformations}. The Weyl form of these transformations is
\begin{equation}
SW(f)S^{+}=W(Uf+Vf^{\ast }).  \label{w9}
\end{equation}
The Weyl relations (and consequently the canonical commutation relations)
are preserved, if the skew symmetric form (\ref{h13}) is invariant against
the $\mathbb{R}$-linear mapping
\begin{equation}
f\in \mathcal{H}_{\mathbb{R}}\rightarrow R(U,V)f:=Uf+Vf^{\ast }\in \mathcal{H%
}_{\mathbb{R}},  \label{w10}
\end{equation}
i.e.
\begin{equation}
\omega (Rf,Rg)\equiv \omega (Uf+Vf^{\ast },Ug+Vg^{\ast })=\omega (f,g).
\label{w11}
\end{equation}
for all $f,g\in \mathcal{H}_{\mathbb{R}}$.
\end{enumerate}

The transformations (\ref{w10}) which satisfy (\ref{w11}) form the \textit{%
symplectic group} of the Hilbert space $\mathcal{H}_{\mathbb{R}}$, and the
transformations (\ref{w9}) generate a unitary ray representation of this
group on the Fock space $\mathcal{S}(\mathcal{H})$. So far we have only
assumed that $U$ and $V$ are bounded linear operators on $\mathcal{H}$ (and
consequently also on $\mathcal{H}_{\mathbb{R}}$). For infinite dimensional
Hilbert spaces $\mathcal{H}$ -- needed for quantum field theory -- an
additional constraint turns out to be necessary: In order to obtain a
unitary ray representation on $\mathcal{S}(\mathcal{H})$ the operator $V$
has to be a Hilbert-Schmidt operator, see \cite{Friedrichs:1953, Shale:1962}.

\section{The symplectic group\label{sympl}}

\subsection{Definition}

In this Section we give a more explicit definition of the symplectic
transformations and recapitulate some identities which are needed for the
subsequent calculations. We identify $\mathcal{H}_{\mathbb{R}}$ with the
diagonal subspace $\mathcal{H}_{diag}\subset \mathcal{H}\times \mathcal{H}%
^{\ast }$, see the beginning of Sect. \ref{Hilbert}. The space $\mathcal{H}%
\times \mathcal{H}^{\ast }$ has elements $\left(
\begin{array}{c}
f \\
g^{\ast }
\end{array}
\right) $ with $f,g\in \mathcal{H}$. On $\mathcal{H}\times \mathcal{H}^{\ast
}$ we define the operators $\Delta =\left(
\begin{array}{cc}
0 & I \\
I & 0
\end{array}
\right) \;$and$\;\hat{M}=\left(
\begin{array}{cc}
I & 0 \\
0 & -I
\end{array}
\right) .$ The matrix array $\hat{R}=\left(
\begin{array}{cc}
U & V \\
X & Y
\end{array}
\right) $ of operators $U,V,X,Y\in \mathcal{L}(\mathcal{H})$ yields a
bounded linear operator on $\mathcal{H}\times \mathcal{H}^{\ast }$%
\begin{equation*}
\left(
\begin{array}{cc}
U & V \\
X & Y
\end{array}
\right) \left(
\begin{array}{c}
f \\
g^{\ast }
\end{array}
\right) =\left(
\begin{array}{c}
Uf+Vg^{\ast } \\
Xf+Yg^{\ast }
\end{array}
\right) .
\end{equation*}

\begin{definition}
\label{symplectic1} The operator $\hat{R}$ is a symplectic transformation,
if it satisfies the constraints
\begin{equation}
\hat{R}\hat{M}\hat{R}^{+}=\hat{M}\;\mathrm{and}\;\Delta \overline{\hat{R}}%
\Delta =\hat{R}.  \label{g2}
\end{equation}
The set of all these transformations is denoted by $\widehat{S}p(\mathcal{H}%
) $.
\end{definition}

The second constraint in (\ref{g2}) implies that $\hat{R}$ has the form
\begin{equation}
\hat{R}=\hat{R}(U,V)=\left(
\begin{array}{cc}
U & V \\
\bar{V} & \bar{U}
\end{array}
\right) .  \label{g4}
\end{equation}
Thereby $\bar{U}$ and $\bar{V}$ are the complex conjugate operators of $U$
and $V$, respectively, as defined in Sect \ref{Hilbert}. The identity
operator $\hat{I}=\left(
\begin{array}{cc}
I & 0 \\
0 & I
\end{array}
\right) $ is an element of $\widehat{S}p(\mathcal{H})$. The product of two
matrix operators (\ref{g4})
\begin{equation}
\hat{R}_{2}\hat{R}_{1}=\left(
\begin{array}{cc}
U_{2} & V_{2} \\
\bar{V}_{2} & \bar{U}_{2}
\end{array}
\right) \left(
\begin{array}{cc}
U_{1} & V_{1} \\
\bar{V}_{1} & \bar{U}_{1}
\end{array}
\right) =\left(
\begin{array}{cc}
U_{2}U_{1}+V_{2}\bar{V}_{1} & U_{2}V_{1}+V_{2}\bar{U}_{1} \\
\bar{V}_{2}U_{1}+\bar{U}_{2}\bar{V}_{1} & \bar{V}_{2}V_{1}+\bar{U}_{2}\bar{U}%
_{1}
\end{array}
\right)  \label{g5}
\end{equation}
is also an element of $\widehat{S}p(\mathcal{H})$. From the first identity
of (\ref{g2}) follows the inverse of $\hat{R}$ as
\begin{equation}
\hat{R}^{-1}=\hat{M}\hat{R}^{+}\hat{M}=\left(
\begin{array}{cc}
U^{+} & -V^{T} \\
-V^{+} & U^{T}
\end{array}
\right) .  \label{g6}
\end{equation}
On the other hand, if a matrix operator (\ref{g4}) satisfies $\hat{R}^{-1}=%
\hat{M}\hat{R}^{+}\hat{M}$, then the conditions of Definition \ref
{symplectic1} apply to $\hat{R}$, and $\hat{R}$ is an element of $\widehat{S}%
p(\mathcal{H})$. From $\hat{R}^{-1}\hat{R}=I$ we have $\hat{R}^{+}\hat{M}%
\hat{R}=\hat{M}\;$such\thinspace that
\begin{equation}
\hat{R}^{+}=\left(
\begin{array}{cc}
U^{+} & V^{T} \\
V^{+} & U^{T}
\end{array}
\right) \in \widehat{S}p(\mathcal{H}).  \label{g7}
\end{equation}
But then also the operator (\ref{g6}) is an element of $\widehat{S}p(%
\mathcal{H})$, and the set $\widehat{S}p(\mathcal{H})$ is a group with
identity $\hat{I}$ and multiplication (\ref{g5}).

The (equivalent) identities $\hat{R}\hat{R}^{-1}=I$ and $\hat{R}^{-1}\hat{R}%
=I$ (with $\hat{R}^{-1}$ given by (\ref{g6})) are satisfied if the following
(again equivalent) conditions hold
\begin{align}
UU^{+}-VV^{+}& =I,\;UV^{T}=VU^{T},  \label{g8} \\
U^{+}U-V^{T}\bar{V}& =I,\;U^{T}\bar{V}=V^{+}U.  \label{g9}
\end{align}
Hence $\left\| U\right\| \geq 1$ and $U$ has an inverse. Then the identities
\begin{equation}
U^{-1}V=V^{T}\left( U^{-1}\right) ^{T},\quad \bar{V}U^{-1}=\left(
U^{-1}\right) ^{T}V^{+}  \label{g10}
\end{equation}
follow. Therefore the operators $U^{-1}V$ and $\bar{V}U^{-1}$ are symmetric.
Moreover we obtain from (\ref{g8}) -- (\ref{g10})
\begin{align}
I-\left( U^{-1}V\right) \left( U^{-1}V\right) ^{+}& =\left( U^{+}U\right)
^{-1},  \label{g11} \\
I-\left( \bar{V}U^{-1}\right) ^{+}\left( \bar{V}U^{-1}\right) & =\left(
UU^{+}\right) ^{-1}.  \label{g12}
\end{align}
\newline
The operator norms of $U^{-1}V$ and $\bar{V}U^{-1}$ therefore satisfy
\begin{equation}
\left\| U^{-1}V\right\| ^{2}=\left\| \bar{V}U^{-1}\right\| ^{2}=1-\left\|
U\right\| ^{-2}<1.  \label{g13}
\end{equation}

The group element $\hat{R}(U,V)\in \widehat{S}p(\mathcal{H})$ maps $\mathcal{%
H}_{diag}\subset \mathcal{H}\times \mathcal{H}^{\ast }$ into itself
\begin{equation}
\hat{R}(U,V)\left(
\begin{array}{c}
f \\
f^{\ast }
\end{array}
\right) =\left(
\begin{array}{cc}
U & V \\
\bar{V} & \bar{U}
\end{array}
\right) \left(
\begin{array}{c}
f \\
f^{\ast }
\end{array}
\right) =\left(
\begin{array}{c}
Uf+Vf^{\ast } \\
\bar{U}f^{\ast }+\bar{V}f
\end{array}
\right) \in \mathcal{H}_{diag}.  \label{g14}
\end{equation}
The operator $\hat{R}(U,V)$ is therefore uniquely determined by the
following $\mathbb{R}$-linear mapping $R(U,V)$ on $\mathcal{H}$ (more
precisely on $\mathcal{H}_{\mathbb{R}}$)
\begin{equation}
R(U,V)f=Uf+Vf^{\ast }.  \label{g15}
\end{equation}
The calculations presented above imply

\begin{lemma}
The skew symmetric form (\ref{h13}) is invariant against the $\mathbb{R}$-
linear mapping $R$ on $\mathcal{H}_{R}$ if and only if $R$ has the form (\ref
{g15}) where $U$ and $V$ are bounded operators on $\mathcal{H}$, which
satisfy the relations (\ref{g8}) and (\ref{g9}).
\end{lemma}

In the sequel we often refer to (\ref{g15}) as the symplectic
transformation. The product and the inverse follow from (\ref{g5}) and (\ref
{g6}) as
\begin{align}
R(U_{2},V_{2})R(U_{1},V_{1})& =R(U_{2}U_{1}+V_{2}\bar{V}_{1},U_{2}V_{1}+V_{2}%
\bar{U}_{1}),  \label{g16} \\
R^{-1}(U,V)& =R(U^{+},-V^{T}).  \label{g17}
\end{align}
The set of these transformations forms the group of symplectic
transformations, which will be denoted by $Sp(\mathcal{H})$. The identity of
the group is $R(I,0)$. In order to derive a unitary representation of this
group on the Fock space $\mathcal{S}(\mathcal{H})$ an additional constraint
is necessary if $\dim \mathcal{H}$ is infinite: The operator $V$ has to be a
Hilbert-Schmidt operator \cite{Friedrichs:1953, Shale:1962}. This constraint
is stable under the group operations (\ref{g16}) and (\ref{g17}).

\begin{definition}
\label{symplectic2}The group $Sp_{2}(\mathcal{H})$ is the subgroup of all
transformations (\ref{g15}) $R(U,V)\in Sp(\mathcal{H})$ with a bounded
operator $U\in \mathcal{L}(\mathcal{H})$ and a Hilbert-Schmidt operator $%
V\in \mathcal{L}_{2}(\mathcal{H})$.
\end{definition}

In \cite{Ottesen:1995} and \cite{Shale:1962} the elements of $Sp_{2}(%
\mathcal{H})$ are called restricted symplectic transformations, in \cite
{Berezin:1966} proper canonical transformations.

Let $U\in \mathcal{L}(\mathcal{H})$ be a unitary operator, then $R(U,0)$ is
an isometric transformation in $Sp_{2}(\mathcal{H}).$ Since $R(U,0)f=Uf$ for
all $f\in \mathcal{H}_{\mathbb{R}}$ we simply write $U$ for this
transformation. In \cite{Shale:1962} Lemma 2.3 it has been derived that any
symplectic transformation $R\in Sp_{2}(\mathcal{H})$ can be factorized in
the form $R=U_{1}DU_{2}$. Thereby $U_{1,2}$ are two unitary transformations,
and $D$ is a real positive transformation in $Sp_{2}(\mathcal{H})$. This
positive transformation has the form $D=R(\cosh A,\sinh A)$ with a real
self-adjoint Hilbert-Schmidt operator $A$ on $\mathcal{H}$.

\subsection{Transformations of the Siegel disc}

There is a non-linear representation of the restricted symplectic group by
transformations on the Siegel disc, investigated by Siegel for the finite
dimensional case \cite{Siegel:1943}. Here we extend some of these results to
the case of infinite dimensions.

From the definition of the Siegel disc follows that a Hilbert-Schmidt
operator $Z=Z^{T}$ is an element of $\mathbf{D}_{1}$ if and only if
\begin{equation}
I-ZZ^{+}>0\quad \mathrm{(all~eigenvalues~strictly~positive)}.  \label{s.1}
\end{equation}

\begin{lemma}
\label{transit}For all $R\in Sp_{2}(\mathcal{H})$ the transformation
\begin{equation}
Z\rightarrow \widetilde{Z}=\zeta (R;Z):=\left( UZ+V\right) \left( \bar{U}+%
\bar{V}Z\right) ^{-1}=\left( U^{+}+ZV^{+}\right) ^{-1}\left(
V^{T}+ZU^{T}\right)  \label{s.2}
\end{equation}
is an automorphism of the set $\mathbf{D}_{1}$. Thereby the group $Sp_{2}(%
\mathcal{H})$ acts transitively on $\mathbf{D}_{1}$.
\end{lemma}

\begin{proof}
For $R\in Sp_{2}(\mathcal{H})$ we have $V\in \mathcal{L}_{2}(\mathcal{H})$,
and $\widetilde{Z}$ is a Hilbert-Schmidt operator. From (\ref{g13}) $\left\|
U^{-1}V\right\| ^{2}=\left\| \bar{V}U^{-1}\right\| ^{2}=1-\left\| U\right\|
^{-2}<1$ and $\left| Z\right| <1$ we know that \newline
$\left\| U^{-1}VZ\right\| <1$, therefore the operator $U+VZ=U\left(
I+U^{-1}VZ\right) $ is invertible. Hence
\begin{equation*}
\begin{array}{l}
I-\widetilde{Z}\widetilde{Z}^{+}=I-\left( U^{+}+ZV^{+}\right) ^{-1}\left(
ZU^{T}+V^{T}\right) \left( \bar{U}Z^{+}+\bar{V}\right) \left(
U+VZ^{+}\right) ^{-1} \\
=\left( U^{+}+ZV^{+}\right) ^{-1}\left\{ \left( U^{+}+ZV^{+}\right) \left(
U+VZ^{+}\right) -\left( ZU^{T}+V^{T}\right) \left( \bar{U}Z^{+}+\bar{V}%
\right) \right\} \left( U+VZ^{+}\right) ^{-1} \\
=\left( U^{+}+ZV^{+}\right) ^{-1}\left\{ I-ZZ^{+}\right\} \left(
U+VZ^{+}\right) ^{-1}>0,
\end{array}
\end{equation*}
since $I-ZZ^{+}>0$. \newline
The proof of the transitivity follows as in the finite dimensional case, see
\cite{Siegel:1943}. Let $Z\in \mathbf{D}_{1}$ then $I-ZZ^{+}>0$ and we can
determine a $U\in \mathcal{L}(\mathcal{H})$ such that $U\left(
I-ZZ^{+}\right) U^{+}=I.$ A special choice is $U=\left( I-ZZ^{+}\right) ^{-%
\frac{1}{2}}\geq I$. The pair $U$ and $V=UZ\in \mathcal{L}_{2}(\mathcal{H})$
satisfies the identities (\ref{g8}), and we easily derive $\zeta
(R;0)=U^{+-1}V^{T}=Z$.
\end{proof}

The mapping $\zeta $ satisfies the rules
\begin{equation}
\begin{array}{l}
\zeta (id;Z)=Z \\
\zeta (R_{2};\zeta (R_{1};Z))=\zeta (R_{2}R_{1};Z).
\end{array}
\label{s5}
\end{equation}
Hence $R\rightarrow \zeta (R;\,.\,)$ is a (nonlinear) representation of the
group $Sp(\mathcal{H})$. If $R=R(U,0)$ with a unitary operator $U$ then $%
\zeta $ has the simple form $\zeta (R;Z)=UZU^{T}$.

\section{Unitary representations of the symplectic group}

A unitary ray representation of $Sp_{2}(\mathcal{H})$ in $\mathcal{S}(%
\mathcal{H})$ has the following properties:
\begin{equation}
\begin{array}{c}
R\in Sp_{2}(\mathcal{H})\longmapsto T\left( R\right) \;\mathrm{%
unitary\,operator\,on}\;\mathcal{S}(\mathcal{H}) \\
T(id)=I,\,T^{-1}\left( R\right) =T^{+}\left( R\right) =T\left( R^{-1}\right)
\\
T(R_{2})T(R_{1})=\chi (R_{2},R_{1})T(R_{2}R_{1}) \\
\mathrm{with}\;\chi (R_{2},R_{1})\in \mathbb{C},\,\left| \chi
(R_{2},R_{1})\right| =1.
\end{array}
\label{r1}
\end{equation}
In this section we construct a unitary ray representation by giving an
explicit formula for $T(R)$ acting on ultracoherent states. As a first step $%
T(R)$ is defined as an isometric operator on the set of exponential vectors
in Sect \ref{ansatz}. This operator can be extended by linearity and
continuity to a unitary operator on the Fock space. In Sect. \ref{ultra} we
derive an explicit formula for the action of $T(R)$ on ultracoherent
vectors. In Sect. \ref{group} we prove that $R\in Sp_{2}(\mathcal{H}%
)\longmapsto T\left( R\right) $ is a ray representation on the linear span
of all ultracoherent vectors. Hence $T\left( R\right) $ is a unitary ray
representation of the restricted symplectic group on $\mathcal{S}(\mathcal{H}%
)$. Finally we prove in Sect. \ref{Bogol} that the operators $T(R)$ are
Bogoliubov transformations, i.e. they generate homogeneous linear canonical
transformations.

\subsection{Representation of the group $Sp_{2}(\mathcal{H})$\label{rep}}

\subsubsection{Ansatz for coherent states\label{ansatz}}

Let $R=R(U,V)$ be a symplectic transformation of the group $Sp_{2}(\mathcal{H%
})$ -- i.e., $U\in \mathcal{L}(\mathcal{H})$ and $V\in \mathcal{L}_{2}(%
\mathcal{H})$ -- then $\left| U\right| :=\sqrt{UU^{+}}=\sqrt{I+VV^{+}}\geq I$
has the property $\left| U\right| -I\in \mathcal{L}_{1}(\mathcal{H})$ and
the determinants $\det \left| U\right| \geq 1$ and $\det \left| U\right|
^{-1}=\left( \det \left| U\right| \right) ^{-1}$ are well defined.

The representation $T(R)$ of the group $Sp_{2}(\mathcal{H})$ is now defined
on the set of exponential vectors by
\begin{equation}
T(R)\exp f:=\left( \det \left| U\right| \right) ^{-\frac{1}{2}}\Phi \left(
U^{+-1}V^{T},\,U^{+-1}f\right) \exp \left( -\frac{1}{2}\left\langle f\mid
V^{+}U^{+-1}f\right\rangle \right) .  \label{r2}
\end{equation}
Since $V$ is a Hilbert-Schmidt operator, the relations (\ref{g10}) and (\ref
{g13}) imply that the mapping $U^{+-1}V^{T}=V\bar{U}^{-1}$ is an element of
the Siegel unit disc. Hence the ultracoherent vector is an element of the
Fock space $\mathcal{S}(\mathcal{H})$. In the special case of a unitary
transformation $R(U,0)$ with $U$ unitary the ansatz (\ref{r2}) has the
simple form $T(R)\exp f=\exp (Uf)$ such that, see (\ref{h7}),
\begin{equation}
T(R(U,0))=\Gamma (U).  \label{r3}
\end{equation}
The transformation $T(R)$ defined in (\ref{r2}) can then be extended by
linearity onto the linear span $\mathcal{S}_{coh}(\mathcal{H})$ of all
exponential vectors. Thereby the identity of the group $R(I_{\mathcal{H}},0)$
is mapped onto the unit operator on $\mathcal{S}_{coh}(\mathcal{H})$. In the
subsequent part of this Section it is shown that $R\in Sp_{2}(\mathcal{H}%
)\rightarrow T(R)$ is actually a unitary ray representation of the group $%
Sp_{2}(\mathcal{H})$ on the Fock space $\mathcal{S}(\mathcal{H})$.

\begin{lemma}
\label{isometric}The operator (\ref{r2}) has a unique extension to a linear
unitary mapping on $\mathcal{S}(\mathcal{H}).$
\end{lemma}

\begin{proof}
For the proof of this statement we calculate the inner product
\begin{equation*}
\begin{array}{c}
\left( T(R)\exp f\mid T(R)\exp g\right) =\det \left| U\right| ^{-1}\exp
\left( -\frac{1}{2}\overline{\left\langle f\mid V^{+}U^{+-1}f\right\rangle }-%
\frac{1}{2}\left\langle g\mid V^{+}U^{+-1}g\right\rangle \right) \\
\times \left( \Phi \left( U^{+-1}V^{T},\,U^{+-1}f\right) \mid \Phi \left(
U^{+-1}V^{T},\,U^{+-1}g\right) \right) .
\end{array}
\end{equation*}
The inner product
\begin{equation*}
\begin{array}{c}
\left( \Phi \left( U^{+-1}V^{T},\,U^{+-1}f\right) \mid \Phi \left(
U^{+-1}V^{T},\,U^{+-1}g\right) \right) =\det \left( I-U^{+-1}V^{T}\bar{V}%
U^{-1}\right) ^{-\frac{1}{2}} \\
\times \exp \left( \frac{1}{2}\left\langle U^{+-1}g\mid
D\;U^{+-1}g\right\rangle +\frac{1}{2}\overline{\left\langle U^{+-1}f\mid
D\;U^{+-1}f\right\rangle }\right) \\
\times \exp \left\langle U^{T-1}f^{\ast }\mid (I-U^{+-1}V^{T}\bar{V}%
U^{-1})^{-1}\;U^{+-1}g\right\rangle ,
\end{array}
\end{equation*}
follows from (\ref{h19}). Thereby $D$ is given by $D=\bar{V}%
U^{-1}(I-U^{+-1}V^{T}\bar{V}U^{-1})^{-1}$. Since \newline
$I-U^{+-1}V^{T}\bar{V}U^{-1}\overset{(\ref{g12})}{=}\left( UU^{+}\right)
^{-1}$ we have $D=\bar{V}U^{+}$ and $\det \left( I-U^{+-1}V^{T}\bar{V}%
U^{-1}\right) =\left( \det \left| U\right| \right) ^{-2}$. We finally obtain
\begin{equation}
\left( T(R)\exp f\mid T(R)\exp g\right) =\left( \exp f\mid \exp g\right)
\label{r4}
\end{equation}
for all $f,g\in \mathcal{H}$. Lemma \ref{isom} then implies that $T(R)$ can
be extended to an isometric mapping, which we denote by the same symbol.

The calculation of $\left( \exp g\mid T^{+}(R)\exp f\right) =\left( T(R)\exp
g\mid \exp f\right) $ using (\ref{h19}) yields
\begin{equation}
T^{+}(R)\exp f=\left( \det \left| U\right| \right) ^{-\frac{1}{2}}\Phi
\left( -U^{-1}V,U^{-1}f\right) \exp \frac{1}{2}\left\langle f\mid \bar{V}%
U^{-1}f\right\rangle .  \label{r5}
\end{equation}
Inserting (\ref{g6}) into (\ref{r2}) we obtain
\begin{equation}
T(R^{-1})=T^{+}(R),  \label{r6}
\end{equation}
first on $\mathcal{S}_{coh}(\mathcal{H})$ and by continuity on $\mathcal{S}(%
\mathcal{H})$. Since $T(R^{-1})$ is isometric the operator $T^{+}(R)$ is
also an isometric mapping. Hence $T(R)$ is unitary.
\end{proof}

\begin{remark}
\label{finite}If $\mathcal{H}$ is finite dimensional, we can use the
determinant $\det U$ (instead of $\det \left| U\right| $) in the definition (%
\ref{r2}).
\end{remark}

\begin{remark}
\label{BF2}The representations of the finite dimensional symplectic group
have been investigated in \cite{Bargmann:1970, Itzykson:1967, KMS:1975}
using the complex wave representation of the Fock space. These authors
calculate the kernel function $\left( \exp g\mid T(R)\exp f\right) $ of the
operator $T(R)$. The ansatz (\ref{r2}) is motivated by these papers.
\end{remark}

\subsubsection{Extension to ultracoherent vectors\label{ultra}}

With help of the relation (\ref{h19}) we can derive a closed formula for $%
T(R)$ operating on ultracoherent vectors
\begin{equation}
\begin{array}{c}
\left( \exp z\mid T(R)\Phi \left( Z,\,f\right) \right) =\left( T^{+}(R)\exp
z\mid \Phi \left( Z,\,f\right) \right) \\
\overset{(\ref{r5})}{=}\left( \det \left| U\right| \right) ^{-\frac{1}{2}%
}\left( \Phi \left( -U^{-1}V,\,U^{-1}z\right) \mid \Phi \left( Z,\,f\right)
\right) \exp \frac{1}{2}\overline{\left\langle z\mid \bar{V}%
U^{-1}z\right\rangle } \\
\overset{(\ref{h19})}{=}\left( \det \left| U\right| \right) ^{-\frac{1}{2}%
}\left( \det (I+ZV^{+}U^{+-1})\right) ^{-\frac{1}{2}}\exp \left( \frac{1}{2}%
\left\langle z^{\ast }\mid V\bar{U}^{-1}z^{\ast }\right\rangle +\frac{1}{2}%
\left\langle \bar{U}^{-1}z^{\ast }\mid C\bar{U}^{-1}z^{\ast }\right\rangle
\right) \\
\times \exp \left( \left\langle \bar{U}^{-1}z^{\ast }\mid
(I+ZV^{+}U^{+-1})^{-1}f\right\rangle +\frac{1}{2}\left\langle f\mid
Df\right\rangle \right)
\end{array}
\label{r8}
\end{equation}
with the operators
\begin{equation}
\begin{array}{c}
C=(I+ZV^{+}U^{+-1})^{-1}Z, \\
D=-V^{+}(U^{+}+ZV^{+})^{-1}.
\end{array}
\label{r9}
\end{equation}
Since \ $U^{+-1}C\bar{U}^{-1}+V\bar{U}%
^{-1}=(U^{+}+ZV^{+})^{-1}(V^{T}+ZU^{T})=\zeta (R;Z)$, we finally obtain
\begin{equation}
\begin{array}{l}
T(R)\,\Phi \left( Z,\,f\right) =\left( \det \left| U\right| \right) ^{-\frac{%
1}{2}}\left( \det (I+ZV^{+}U^{+-1})\right) ^{-\frac{1}{2}} \\
\times \Phi \left( \zeta (R;Z),\,(U^{+}+ZV^{+})^{-1}f\right) \exp \left( -%
\frac{1}{2}\left\langle f\mid V^{+}(U^{+}+ZV^{+})^{-1}f\right\rangle \right)
.
\end{array}
\label{r10}
\end{equation}

\subsubsection{The multiplication law\label{group}}

In the next step we prove
\begin{equation}
T(R_{2})T(R_{1})=\chi (R_{2},R_{1})T(R_{3})\quad \mathrm{if}%
\;R_{2}R_{1}=R_{3}  \label{r11}
\end{equation}
with a multiplier $\chi (R_{2},R_{1})\in \mathbb{C},\,\left| \chi
(R_{2},R_{1})\right| =1$.

Let
\begin{equation}
\begin{array}{l}
Z_{1}=\zeta (R_{1};Z)=(U_{1}^{+}+ZV_{1}^{+})^{-1}(V_{1}^{T}+ZU_{1}^{T}) \\
Z_{2}=\zeta (R_{2};Z_{1})=\zeta (R_{3};Z)
\end{array}
\label{r12}
\end{equation}
see (\ref{s5}), then we obtain from (\ref{r10})
\begin{equation*}
\begin{array}{l}
T(R_{1})\Phi \left( Z,\,f\right) =\left( \det \left| U_{1}\right| \right) ^{-%
\frac{1}{2}}\left( \det (I+ZV_{1}^{+}U_{1}^{+-1})\right) ^{-\frac{1}{2}} \\
\times \Phi \left( Z_{1},(U_{1}^{+}+ZV_{1}^{+})^{-1}f\right) \exp \left( -%
\frac{1}{2}\left\langle f\mid
V_{1}^{+}(U_{1}^{+}+ZV_{1}^{+})^{-1}f\right\rangle \right)
\end{array}
\end{equation*}
and
\begin{equation}
\begin{array}{l}
T(R_{2})T(R_{1})\Phi \left( Z,\,f\right) \\
=\left( \det \left| U_{2}\right| \right) ^{-\frac{1}{2}}\left( \det
(I+Z_{1}V_{2}^{+}U_{2}^{+-1})\right) ^{-\frac{1}{2}}\left( \det \left|
U_{1}\right| \right) ^{-\frac{1}{2}}\left( \det
(I+ZV_{1}^{+}U_{1}^{+-1})\right) ^{-\frac{1}{2}} \\
\times \Phi \left(
Z_{2},\,(U_{2}^{+}+Z_{1}V_{2}^{+})^{-1}(U_{1}^{+}+ZV_{1}^{+})^{-1}f\right)
\\
\times \exp \left( -\frac{1}{2}\left\langle (U_{1}^{+}+ZV_{1}^{+})^{-1}f\mid
V_{2}^{+}(U_{2}^{+}+Z_{1}V_{2}^{+})^{-1}(U_{1}^{+}+ZV_{1}^{+})^{-1}f\right%
\rangle \right) \\
\times \exp \left( -\frac{1}{2}\left\langle f\mid
V_{1}^{+}(U_{1}^{+}+ZV_{1}^{+})^{-1}f\right\rangle \right) .
\end{array}
\label{r13}
\end{equation}
The tensor of second degree in the exponent immediately follows as $%
Z_{2}=\zeta (R_{2};Z_{1})=\zeta (R_{3};Z)$. The operator product $%
(U_{2}^{+}+Z_{1}V_{2}^{+})^{-1}(U_{1}^{+}+ZV_{1}^{+})^{-1}$ is calculated
with
\begin{equation*}
\begin{array}{l}
U_{2}^{+}+Z_{1}V_{2}^{+}=(U_{1}^{+}+ZV_{1}^{+})^{-1}\left(
(U_{1}^{+}+ZV_{1}^{+})U_{2}^{+}+(V_{1}^{T}+ZU_{1}^{T})V_{2}^{+}\right) \\
=(U_{1}^{+}+ZV_{1}^{+})^{-1}\left(
U_{1}^{+}U_{2}^{+}+ZV_{1}^{+}U_{2}^{+}+V_{1}^{T}V_{2}^{+}+ZU_{1}^{T}V_{2}^{+}\right) =(U_{1}^{+}+ZV_{1}^{+})^{-1}(U_{3}^{+}+ZV_{3}^{+})
\end{array}
\end{equation*}
as
\begin{equation}
(U_{2}^{+}+Z_{1}V_{2}^{+})^{-1}(U_{1}^{+}+ZV_{1}^{+})^{-1}=(U_{3}^{+}+ZV_{3}^{+})^{-1}.
\label{r20}
\end{equation}
Hence we obtain
\begin{align}
T(R_{2})T(R_{1})\Phi \left( Z,\,f\right) & =\chi \,T(R_{3})\Phi \left(
Z,\,f\right)  \notag \\
& \times \exp \left( -\frac{1}{2}\left( \alpha _{12}-\left\langle f\mid
V_{3}^{+}(U_{3}^{+}+ZV_{3}^{+})^{-1}f\right\rangle \right) \right)
\label{r21}
\end{align}
with
\begin{equation}
\chi =\sqrt{\frac{\det \left| U_{3}\right| }{\det \left| U_{1}\right| \det
\left| U_{2}\right| }}\sqrt{\frac{\det (I+ZV_{3}^{+}U_{3}^{+-1})}{\det
(I+ZV_{1}^{+}U_{1}^{+-1})\det (I+Z_{1}V_{2}^{+}U_{2}^{+-1})}}  \label{r22}
\end{equation}
and
\begin{equation}
\begin{array}{c}
\alpha _{12}=\left\langle (U_{1}^{+}+ZV_{1}^{+})^{-1}f\mid
V_{2}^{+}(U_{2}^{+}+Z_{1}V_{2}^{+})^{-1}(U_{1}^{+}+ZV_{1}^{+})^{-1}f\right%
\rangle \\
+\left\langle f\mid V_{1}^{+}(U_{1}^{+}+ZV_{1}^{+})^{-1}f\right\rangle .
\end{array}
\label{r23}
\end{equation}

Now we choose $Z=0$. Then (\ref{r22}) simplifies to
\begin{equation}
\chi (R_{2},R_{1})=\sqrt{\frac{\det \left| U_{3}\right| }{\det \left|
U_{1}\right| \det \left| U_{2}\right| \det \left(
U_{1}^{+-1}U_{3}^{+}U_{2}^{+-1}\right) }},  \label{r24}
\end{equation}
which depends only on the group elements. In the next step we evaluate (\ref
{r23}) for $Z=0$
\begin{equation*}
\begin{array}{c}
\alpha _{12}=\left\langle U_{1}^{+-1}f\mid
V_{2}^{+}(U_{2}^{+}+U_{1}^{+-1}V_{1}^{T}V_{2}^{+})^{-1}U_{1}^{+-1}f\right%
\rangle +\left\langle f\mid V_{1}^{+}U_{1}^{+-1}f\right\rangle \\
=\left\langle f\mid \left( \bar{U}%
_{1}{}^{-1}V_{2}^{+}(U_{2}^{+}+U_{1}^{+-1}V_{1}^{T}V_{2}^{+})^{-1}U_{1}^{+-1}+V_{1}^{+}U_{1}^{+-1}\right) f\right\rangle .
\end{array}
\end{equation*}
Since $(U_{2}^{+}+U_{1}^{+-1}V_{1}^{T}V_{2})^{-1}=\left(
U_{1}^{+}U_{2}^{+}+V_{1}^{T}V_{2}\right) ^{-1}U_{1}^{+}$ we have
\begin{equation*}
\begin{array}{c}
\bar{U}%
_{1}^{-1}V_{2}^{+}(U_{2}^{+}+U_{1}^{+-1}V_{1}^{T}V_{2}^{+})^{-1}U_{1}^{+-1}+V_{1}^{+}U_{1}^{+-1}
\\
=\left( V_{1}^{+}U_{1}^{+-1}\left(
U_{1}^{+}U_{2}^{+}+V_{1}^{T}V_{2}^{+}\right) +\bar{U}_{1}^{-1}V_{2}^{+}%
\right) \left( U_{1}^{+}U_{2}^{+}+V_{1}^{T}V_{2}^{+}\right) ^{-1} \\
\overset{(\ref{g5})(\ref{g10})}{=}\left( V_{1}^{+}U_{2}^{+}+\bar{U}_{1}^{-1}(%
\bar{V}_{1}V_{1}^{T}+I)V_{2}^{+}\right) U_{3}^{+-1} \\
\overset{(\ref{g8})}{=}\left( V_{1}^{+}U_{2}^{+}+\bar{U}_{1}^{-1}\bar{U}%
_{1}U_{1}^{T}V_{2}^{+}\right) U_{3}^{+-1}=\left(
V_{1}^{+}U_{2}^{+}+U_{1}^{T}V_{2}^{+}\right) U_{3}^{+-1}\overset{(\ref{g5})}{%
=}V_{3}^{+}U_{3}^{+-1},
\end{array}
\end{equation*}
such that
\begin{equation}
\alpha _{12}=\left\langle f\mid V_{3}^{+}U_{3}^{+}{}^{-1}f\right\rangle .
\label{r26}
\end{equation}

The identity (\ref{r21}) together with (\ref{r24}) and (\ref{r26}) imply
\begin{equation}
T(R_{2})T(R_{1})\exp f=\chi (R_{1},R_{2})\,T(R_{3})\exp f  \label{r27}
\end{equation}
for all $f\in \mathcal{H}$. But then (\ref{r11}) is true as an operator
identity. Since we already know that the operators $T(R)$ are unitary, the
modulus of $\chi $ is $\left| \chi (R_{1},R_{2})\right| =1$.

\begin{remark}
\label{rem1}The identity (\ref{r6}) implies $\chi (R,R^{-1})=\chi
(R^{-1},R)=1$.
\end{remark}

\begin{remark}
\label{rem2}Since $\det \left| KUK^{-1}\right| =\det \left| U\right| $ the
definition (\ref{r2}) implies that
\begin{equation}
T(KRK^{-1})=\Gamma (K)T(R)\Gamma (K^{+})  \label{w28}
\end{equation}
holds for all unitary transformations $K$ and all $R\in Sp_{2}(\mathcal{H})$
without additional phase factor.
\end{remark}

\begin{remark}
If $\mathcal{H}$ is finite dimensional, the determinant $\det U$ (instead of
$\det \left| U\right| $) can be used in the definition (\ref{r2}). Then the
multiplier $\chi (R_{1},R_{2})$ takes only the values $\pm 1$, see \cite
{Bargmann:1970}.
\end{remark}

\subsection{Weyl operators and Bogoliubov transformations\label{Bogol}}

In this Section we prove that the operators $T(R)$ are Bogoliubov
transformations as defined in Sect.\ref{canonical}.

Given $f,g\in \mathcal{H}$ and $R(U,V)\in Sp_{2}(\mathcal{H})$ we first
calculate
\begin{equation}
\begin{array}{l}
T(R)W(f)\exp g\overset{(\ref{w1})}{=}\exp \left( -\left\langle f^{\ast }\mid
g\right\rangle -\frac{1}{2}\left\| f\right\| ^{2}\right) T(R)\exp (f+g) \\
=\left( \det \left| U\right| \right) ^{-\frac{1}{2}}\Phi \left(
U^{+-1}V^{T},\,U^{+-1}(f+g)\right) \\
\times \exp \left( -\left\langle f^{\ast }\mid g\right\rangle -\frac{1}{2}%
\left\| f\right\| ^{2}-\frac{1}{2}\left\langle f+g\mid
V^{+}U^{+-1}(f+g)\right\rangle \right) .
\end{array}
\label{w33}
\end{equation}
On the other hand $W(Rf)T(R)\exp g$ follows as
\begin{equation}
\begin{array}{l}
W(Rf)T(R)\exp g\overset{(\ref{r2})}{=}\det \left| U\right| ^{-\frac{1}{2}%
}\exp \left( -\frac{1}{2}\left\langle g\mid V\bar{U}^{-1}g\right\rangle
\right) \\
\quad \times W(Uf+Vf^{\ast })\Phi \left( V\bar{U}^{-1},\,U^{+-1}g\right) \\
\overset{(\ref{w4})}{=}\left( \det \left| U\right| \right) ^{-\frac{1}{2}%
}\exp \left( -\frac{1}{2}\left\langle g\mid V^{+}U^{+-1}g\right\rangle
-\left\langle \bar{U}f^{\ast }+\bar{V}f\mid U^{+-1}g\right\rangle \right) \\
\quad \times \exp \left( +\frac{1}{2}\left\langle \bar{U}f^{\ast }+\bar{V}%
f\mid V\bar{U}^{-1}(\bar{U}f^{\ast }+\bar{V}f)-(Uf+Vf^{\ast })\right\rangle
\right) \\
\quad \times \Phi \left( U^{+-1}V^{T},\,Uf+Vf^{\ast }+U^{+-1}g-V\bar{U}^{-1}(%
\bar{U}f^{\ast }+\bar{V}f)\right) \\
=\left( \det \left| U\right| \right) ^{-\frac{1}{2}}\Phi \left(
U^{+-1}V^{T},\,U^{+-1}(f+g)\right) \\
\times \exp (-\left\langle f^{\ast }\mid g\right\rangle -\frac{1}{2}\left\|
f\right\| ^{2}-\frac{1}{2}\left\langle f+g\mid
V^{+}U^{+-1}(f+g)\right\rangle ).
\end{array}
\label{w34}
\end{equation}
The last identity follows using (\ref{g10}) and (\ref{g11}). Hence we have
derived the identity \newline
$T(R)W(f)=W(Rf)T(R)$ on $\mathcal{S}_{coh}(\mathcal{H}).$ But then (\ref{w9}%
) is true on $\mathcal{S}(\mathcal{H})$ for all homogeneous canonical
transformations $S=T(R),\,R\in Sp_{2}(\mathcal{H}).$

\section{Concluding remarks\label{concl}}

In this paper we have given a self-contained construction of the unitary
representation of the (in)finite dimensional symplectic group on the Fock
space. The operators are first defined on the linear span of all exponential
vectors (coherent states) and then extended onto the minimal linear set,
which is closed under the action of Weyl operators and of homogeneous linear
canonical transformations. Actually, the calculations of the paper show that
any ultracoherent vector $\Phi (A,f)$ has a representation $\Phi
(A,f)=const\cdot W(h)T(R)\,1_{vac}$ where $1_{vac}$ is the vacuum state of
the Fock space, $W(h)$ is a Weyl operator and $T(R)$ is a Bogoliubov
transformation. This follows from the transitivity of the action of $Sp_{2}(%
\mathcal{H})$ on the Siegel disc, see Lemma \ref{transit}, and from the
formulas (\ref{w4}) and (\ref{id6}). The states $T(R)\,1_{vac}$ are the
squeezed vacua of quantum optics. Our presentation of canonical
transformations is therefore useful for applications in quantum optics and
in quantum computation \cite{Banerjee/Kupsch:2005, Ken:1988, Kim:1993}.

The constructions given in Sect. \ref{rep} are independent from the explicit
representation of the Fock space. Nevertheless we would like to mention two
of these representations:

\begin{enumerate}
\item  The complex wave representation or Bargmann-Fock representation uses
a reproducing kernel Hilbert space, see Remark \ref{BF}. The operators $T(R)$
have integral kernels $\left( \exp g\mid T(R)\exp f\right) $ also in the
infinite dimensional case. These kernels can easily be calculated from (\ref
{r2}). They differ from the kernels for the finite dimensional group as
given in \cite{Bargmann:1970, Itzykson:1967, KMS:1975} only by the choice of
the determinant, see Remark \ref{finite}.

\item  The real wave representation or Wiener-Segal representation, see \cite
{BSZ:1992, Nielsen:1991}, is closely related to the self-adjoint canonical
field and momentum variables. To transfer the method of Sect \ref{rep} to
this representation one has to use the real form of the symplectic group,
and the Siegel disc has to be substituted by the Siegel upper half plane
(operators of $\mathcal{L}_{2sym}$ with a positive imaginary part). The
ultracoherent vectors are Gaussian functions in the position variable. Also
in this case it is possible to implement Bogoliubov transformations by
integral transforms \cite{Kupsch/Smolyanov:2000a, Kupsch/Smolyanov:2000b}.
\end{enumerate}

The explicit construction of $T(R)\;$on a dense domain of definition -- in
this case on ultracoherent vectors -- has some advantage for the
investigation of one-parameter subgroups and their generators. In the
infinite dimensional case the group $Sp_{2}(\mathcal{H})$ is still a
topological group \cite{Shale:1962}, but the one-parameter subgroups may
have unbounded generators. A Lie group structure has been imposed only under
the additional restriction to bounded generators, see Chapter 3 of \cite
{Ottesen:1995}. For applications in quantum field theory one would like to
consider subgroups with unbounded generators. Take as example a free field
with an unbounded positive one-particle Hamiltonian $M=\bar{M}$. The domain
of definition of $M$ is the dense set $\mathcal{D}(M)\subset \mathcal{H}$.
With $U_{0}(t)=\exp (-iMt)$ we denote the corresponding unitary group on $%
\mathcal{H}$. The linear set $\mathcal{D}_{F}=span\left\{ \exp f\mid f\in
\mathcal{D}(M)\right\} $ is dense in $\mathcal{S}(\mathcal{H})$, and the
free field Hamiltonian $H_{F}=d\Gamma (M)$ is defined on $\mathcal{D}_{F}$
by $d\Gamma (M)\exp f=Mf\vee \exp f$. The unitary group $\exp
(-iH_{F}t)=\Gamma (U_{0}(t))$ is a one parameter subgroup of proper
canonical transformations. If we now submit the free field to a homogeneous
canonical transformation $S=T(R_{1})$ with $R_{1}=R(U_{1},V_{1})\in Sp_{2}(%
\mathcal{H})$, then the resulting unitary group in $\mathcal{S}(\mathcal{H})$
is $S\exp (-iH_{F}t)S^{-1}=T(R_{2}(t))$ with $%
R_{2}(t)=R(U_{1}U_{0}(t)U_{1}^{+}-V_{1}U_{0}(-t)V_{1}^{+},-U_{1}U_{0}(t)V_{1}^{T}+V_{1}U_{0}(-t)U_{1}^{T})
$. The generator of this group is the Hamiltonian of the transformed free
field. Formal differentiation of $T(R_{2}(t))$ to obtain the Hamiltonian is
misleading, as expressions like $U_{1}MU_{1}^{+}+V_{1}MV_{1}^{+}$ might not
be defined because $M$ is unbounded. But our equation (\ref{r2}) gives an
explicit formula for a domain of analytic vectors for this Hamiltonian
within the set of ultracoherent vectors: $\mathcal{D}=T(R_{1})\mathcal{D}%
_{F} $.

A related application is the investigation of the representation of general
one-parameter subgroups of the symplectic group. The complexity of this
problem can be seen from the paper \cite{Ito/Hiroshima:2004}, where results
on the basis of the white noise calculus have been derived.

\appendix

\section{Calculations for ultracoherent vectors\label{uv}}

\subsection{Norm estimates for the tensor product\label{norm}}

The symmetric tensor product is well defined on the algebra $\mathcal{S}%
_{fin}(\mathcal{H})$. To extend it to a larger space we introduce a family
of Hilbert norms \cite{Kupsch/Smolyanov:2000}. For Let $F=\sum_{n}F_{n}\in
\mathcal{S}_{fin}(\mathcal{H})$ with $F_{n}\in \widehat{\mathcal{H}^{\vee n}}
$ then the norm $\left\| F\right\| _{(\alpha )},\,\alpha >0$, is defined by
\begin{equation}
\left\| F\right\| _{(\alpha )}^{2}=\sum_{n}\alpha ^{-2n}\left\|
F_{n}\right\| _{n}^{2}\,.  \label{n1}
\end{equation}
Thereby $\left\| \,.\,\right\| _{n}$ is the Hilbert norm of $\widehat{%
\mathcal{H}^{\vee n}}$ introduced in Sect. \ref{Hilbert}. The completion of $%
\mathcal{S}_{fin}(\mathcal{H})$ with respect to the norm (\ref{n1}) is
called $\mathcal{S}^{(\alpha )}(\mathcal{H})$. We obviously have $\mathcal{S}%
^{(1)}(\mathcal{H})=\mathcal{S}(\mathcal{H})$ and
\begin{equation}
\begin{array}{l}
\left\| F\right\| =\left\| F\right\| _{(1)}\leq \left\| F\right\| _{(\alpha
)}\;\mathrm{if}\;\alpha \in \left( 0,1\right] , \\
\left\| F_{n}\right\| _{n}\leq \alpha ^{n}\left\| F\right\| _{(\alpha )}\;%
\mathrm{if}\;\alpha >0.
\end{array}
\label{n1a}
\end{equation}

\begin{lemma}
\label{Focknorms} The following statements are true:

\begin{enumerate}
\item  For $\alpha \in \left( 0,1\right] $ the algebras $\mathcal{S}_{fin}(%
\mathcal{H})$ and $\mathcal{S}_{coh}(\mathcal{H})$ are dense linear subsets
of $\mathcal{S}^{(\alpha )}(\mathcal{H})$, and $\mathcal{S}^{(\alpha )}(%
\mathcal{H})$ is a dense linear subset of $\mathcal{S}(\mathcal{H})$.

\item  Let $A\in \mathbf{D}_{1}$ then $\exp \Omega (A)\in \mathcal{S}%
^{(\alpha )}(\mathcal{H})$ with $\alpha \in \left( \left\| A\right\| ,1%
\right] $.

\item  Let $0<\alpha ,\beta ,\gamma <1$ with $\alpha +\beta <\gamma \leq 1$,
then there exists a constant $c_{\alpha \beta \gamma }$ such that
\begin{equation}
\left\| F\vee G\right\| _{(\gamma )}\leq c_{\alpha \beta \gamma }\left\|
F\right\| _{(\alpha )}\left\| G\right\| _{(\beta )}  \label{n2}
\end{equation}
for all $F,\,G\in \mathcal{S}_{fin}(\mathcal{H})$.
\end{enumerate}
\end{lemma}

\begin{proof}
1) The norm (\ref{n1}) of an exponential vector $\exp f,\,f\in \mathcal{H}$,
is $\left\| \exp f\right\| _{(\alpha )}^{2}=\left\| \exp (\alpha
^{-1}f)\right\| ^{2}$. Hence $\left\| \exp f\right\| _{(\alpha )}<\infty $
for all $\alpha >0$ and $\mathcal{S}_{coh}(\mathcal{H})$ is a subset of $%
\mathcal{S}^{(\alpha )}(\mathcal{H}),\,\alpha \in \left( 0,1\right] .$ Since
$\mathcal{S}_{coh}(\mathcal{H})$ is dense in $\mathcal{S}(\mathcal{H})$, the
identity (\ref{n1}) implies that $\mathcal{S}_{coh}(\mathcal{H})$ is dense
in $\mathcal{S}^{(\alpha )}(\mathcal{H})$ in the topology (\ref{n1}). The
other statements of 1) are obvious.

2) Assume $A\in \mathbf{D}_{1}$ then for $0<\lambda <\left\| A\right\| ^{-1}$
also $\lambda A$ is an element of $\mathbf{D}_{1}$. Hence the series $\exp
\Omega (\lambda A)=\sum_{n}\frac{\lambda ^{n}}{n!}\left( \Omega (A)\right)
^{\vee n}$ converges within $\mathcal{S}(\mathcal{H})$ and $\left\| \frac{%
\lambda ^{n}}{n!}\left( \Omega (A)\right) ^{\vee n}\right\| _{2n}^{2}\leq
\left\| \exp \Omega (\lambda A)\right\| ^{2}=C<\infty $ follows. The
estimate $\left\| \frac{1}{n!}\left( \Omega (A)\right) ^{\vee n}\right\|
_{2n}^{2}\leq C\lambda ^{-2n}$ implies \newline
$\left\| \exp \Omega (A)\right\| _{(\alpha )}^{2}=\sum_{n}\alpha
^{-2n}\left\| \frac{1}{n!}\left( \Omega (A)\right) ^{\vee n}\right\|
_{2n}^{2}\leq C\sum_{n}(\alpha \lambda )^{-2n}<\infty $ for all $\alpha
>\lambda ^{-1}>\left\| A\right\| $.

3) Let $F=\sum_{n}F_{n}$ and $G=\sum_{n}G_{n}$ be two elements of $\mathcal{S%
}_{fin}(\mathcal{H})$ with $F_{n},G_{n}\in \widehat{\mathcal{H}^{\vee n}}$,
then (\ref{n1a}) implies $\left\| F_{n}\right\| _{n}\leq \alpha ^{n}\left\|
F\right\| _{(\alpha )}$ and $\left\| G_{n}\right\| _{n}\leq \beta
^{n}\left\| G\right\| _{(\beta )}$ with arbitrary $\alpha ,\beta >0$. The
symmetric tensor product $F\vee G$ is then calculated as $F\vee
G=\sum_{n}H_{n}$ with $H_{n}=\sum_{m=0}^{n}F_{m}\vee G_{n-m}\in \widehat{%
\mathcal{H}^{\vee n}}$. From (\ref{h3a}) and (\ref{n1a}) we have
\begin{equation*}
\begin{array}{c}
\left\| \sum_{m=0}^{n}F_{m}\vee G_{n-m}\right\| _{n}\leq \sum_{m=0}^{n}\sqrt{%
\left(
\begin{array}{c}
n \\
m
\end{array}
\right) }\left\| F_{m}\right\| _{m}\left\| G_{n-m}\right\| _{n-m} \\
\leq \left\| F\right\| _{(\alpha )}\left\| G\right\| _{(\beta
)}\sum_{m=0}^{n}\left(
\begin{array}{c}
n \\
m
\end{array}
\right) \alpha ^{m}\beta ^{n-m}=\left\| F\right\| _{(\alpha )}\left\|
G\right\| _{(\beta )}(\alpha +\beta )^{n}.
\end{array}
\end{equation*}
Hence $\left\| F\vee G\right\| _{(\gamma )}^{2}\leq \left\| F\right\|
_{(\alpha )}^{2}\left\| G\right\| _{(\beta )}^{2}\sum_{n}\gamma
^{-2n}(\alpha +\beta )^{2n}$, and the upper bound (\ref{n2}) follows for all
$\alpha ,\beta ,\gamma >0$ with $\alpha +\beta <\gamma \leq 1$.
\end{proof}

The third statement of this Lemma implies

\begin{corollary}
\label{product} Let $\alpha ,\,\beta ,\,\gamma $ be strictly positive
numbers with $\alpha +\beta <\gamma \leq 1$, then the symmetric tensor
product $F,\,G\rightarrow F\vee G=G\vee F$ is a $\mathbb{C}$-bilinear
continuous mapping from $\mathcal{S}^{(\alpha )}(\mathcal{H})\times \mathcal{%
S}^{(\beta )}(\mathcal{H})$ into $\mathcal{S}^{(\gamma )}(\mathcal{H})$.
\end{corollary}

If $A\in \mathbf{D}_{1}$ then we know from the second statement of Lemma \ref
{Focknorms} that $\exp \Omega (A)\in \mathcal{S}^{(\alpha )}(\mathcal{H})$
for some $\alpha \in \left( \left\| A\right\| ,1\right) .$ On the other hand
the exponential vector $\exp f,\,f\in \mathcal{H}$, is an element of $%
\mathcal{S}^{(\beta )}(\mathcal{H})$ with $0<\beta <1-\alpha $. The
Corollary then implies that the symmetric tensor product $\exp \Omega
(A)\vee \exp f=\exp f\vee \exp \Omega (A)$ is an element of the Fock space,
which depends continuously on its factors. The ultracoherent vectors (\ref
{h18}) are therefore defined as elements of $\mathcal{S}(\mathcal{H})$ for $%
A\in \mathbf{D}_{1}$ and $f\in \mathcal{H}$. Moreover, since $f\in \mathcal{H%
}\rightarrow \exp f$ and $A\in \mathbf{D}_{1}\rightarrow \exp \Omega (A)$
are holomorphic functions, the ultracoherent vector (\ref{h18}) $\Phi
(A,f)=\exp \Omega (A)\vee \exp f$ is an analytic function of $A\in \mathbf{D}%
_{1}$ and $f\in \mathcal{H}$.

\subsection{Identities\label{identities}}

Let $F$ and $G$ be two elements of $\mathcal{S}_{coh}(\mathcal{H})$, then
the identity
\begin{equation}
\left( \exp z\mid F\vee G\right) =\left( \exp z\mid F\right) \left( \exp
z\mid G\right)  \label{id1}
\end{equation}
follows. For the proof it is sufficient to choose the exponential vectors $%
F=\exp f$ and $G=\exp g$. This identity can be extended by continuity to $%
F\in \mathcal{S}^{(\alpha )}(\mathcal{H})$ and $G\in \mathcal{S}^{(\beta )}(%
\mathcal{H})$ if $\alpha ,\beta >0$ and $\alpha +\beta <1$.

The inner product $\left( \exp z\mid \exp \Omega (A)\right) $ is calculated
by evaluation of the power series using the product rule (\ref{id1}) as
\begin{equation}
\left( \exp z\mid \exp \Omega (A)\right) =\exp \frac{1}{2}\left( z\vee z\mid
\Omega (A)\right) =\exp \frac{1}{2}\left\langle z^{\ast }\mid Az^{\ast
}\right\rangle .  \label{id2}
\end{equation}
Further application of (\ref{id1}) yields
\begin{equation}
\left( \exp z\mid \exp \Phi (A,f)\right) =\left( \exp z\mid \exp \Omega
(A)\vee \exp f\right) =\exp \left( \frac{1}{2}\left\langle z^{\ast }\mid
Az^{\ast }\right\rangle +\left\langle z^{\ast }\mid f\right\rangle \right) .
\label{id3}
\end{equation}

The inner product (\ref{id3}) can be used to determine the action of the
Weyl operator $W(h)$ on the ultracoherent vector $\Phi (A,f)$. We have
\begin{equation*}
\begin{array}{l}
\left( \exp z\mid W(h)\Phi (A,f)\right) =\left( W(-h)\exp z\mid \Phi
(A,f)\right) \\
\overset{(\ref{w1})}{=}\mathrm{e}^{-\frac{1}{2}\left\| h\right\| ^{2}+\frac{1%
}{2}\left\langle h^{\ast }\mid Ah^{\ast }\right\rangle -\left\langle h^{\ast
}\mid f\right\rangle }\mathrm{e}^{\left\langle f+h-Ah^{\ast }\mid z^{\ast
}\right\rangle +\frac{1}{2}\left\langle z^{\ast }\mid Az^{\ast
}\right\rangle } \\
=\mathrm{e}^{-\frac{1}{2}\left\| h\right\| ^{2}+\frac{1}{2}\left\langle
h^{\ast }\mid Ah^{\ast }\right\rangle -\left\langle h^{\ast }\mid
f\right\rangle }\left( \exp z\mid \Phi (A,f+h-Ah^{\ast })\right)
\end{array}
\end{equation*}
which implies the relation (\ref{w4}).

The restriction of (\ref{w4}) to $\Phi (A,0)$ is
\begin{equation}
W(h)\exp \Omega (A)=\mathrm{e}^{\frac{1}{2}\left\langle h^{\ast }\mid
Ah^{\ast }-h\right\rangle }\Phi (A,h-Ah^{\ast }).  \label{id4}
\end{equation}
Since the norm of $\exp \Omega (A)$ is known and $W(h)$ is unitary, we use (%
\ref{id4}) to calculate the norm of the exponential vectors.

Any operator $A\in \mathbf{D}_{1}$ satisfies $\left\| A\right\| <1$,
therefore $(I-A^{+}A)^{-1}$ and $(I-AA^{+})^{-1}$ are bounded operators. Let
$f$ be a vector in $\mathcal{H}$, then
\begin{equation}
h=(I-AA^{+})^{-1}f+A(I-A^{+}A)^{-1}f^{\ast }  \label{id5}
\end{equation}
is an element of $\mathcal{H}$, which satisfies $h-Ah^{\ast }=f$. With this
vector $h$ we obtain from (\ref{id4})
\begin{equation}
W(h)\exp \Omega (A)=\Phi (A,f)\exp \left( \frac{1}{2}\left\langle f\mid
(I-A^{+}A)^{-1}f^{\ast }+A^{+}(I-AA^{+})^{-1}f\right\rangle \right) .
\label{id6}
\end{equation}
Since the Weyl operator is unitary, we know from (\ref{h17})
\begin{equation}
\left( W(h)\exp \Omega (A)\mid W(h)\exp \Omega (A)\right) =\left( \exp
\Omega (A)\mid \exp \Omega (A)\right) =\det \left( I-A^{+}A\right) ^{-\frac{1%
}{2}}.  \label{id7}
\end{equation}
Substituting (\ref{id6}) into this identity we obtain
\begin{equation}
\begin{array}{l}
\left( \Phi (A,f)\mid \Phi (A,f)\right) =\det \left( I-A^{+}A\right) ^{-%
\frac{1}{2}} \\
\times \exp \left( -\frac{1}{2}\left\langle f^{\ast }\mid
A(I-A^{+}A)^{-1}f^{\ast }\right\rangle +\left\langle f^{\ast }\mid
(I-AA^{+})^{-1}\;f\right\rangle +\frac{1}{2}\left\langle f\mid
A^{+}(I-AA^{+})^{-1}\;f\right\rangle \right) .
\end{array}
\label{id8}
\end{equation}
The inner product $\varphi (A,f;B,g):=\left( \Phi (A,f)\mid \Phi
(B,g)\right) $ of the exponential vectors $\Phi (A,f)$ and $\Phi (B,g)$ is
analytic in $B$ and $g$, and it is antianalytic in $A$ and in $f$. The
function $\varphi (A,f;B,g)$ is uniquely determined by its values on the
diagonals $A=B$ and $f=g$. From (\ref{id8}) we then obtain
\begin{equation*}
\begin{array}{l}
\varphi (A,f;B,g)=\det \left( I-A^{+}B\right) ^{-\frac{1}{2}} \\
\times \exp \left( -\frac{1}{2}\left\langle f^{\ast }\mid C\,f^{\ast
}\right\rangle +\left\langle f^{\ast }\mid (I-BA^{+})^{-1}\;g\right\rangle +%
\frac{1}{2}\left\langle g\mid D\;g\right\rangle \right) ,
\end{array}
\end{equation*}
where the operators $C$ and $D$ are defined in (\ref{h20}). Hence we have
derived the inner product (\ref{h19}) of two ultracoherent vectors.

\begin{remark}
With the diagonalization technique of \cite{KMTP:1967} one can derive that
\newline
$\lim_{J\rightarrow \infty }\sum_{j=0}^{J}\frac{1}{j!}\left( \Omega
(A)+f\right) ^{\vee j}$ is norm convergent for $A\in \mathbf{D}_{1}$ and $%
f\in \mathcal{H}$. We denote the limit by $\exp \left( \Omega (A)+f\right) $%
. Since $\sum_{k=0}^{\infty }\frac{1}{k!}\left( \exp z\mid \left( \Omega
(A)+f\right) ^{\vee k}\right) \newline
=\sum_{m,n=0}^{\infty }\frac{1}{m!n!}\left( \exp z\mid \left( \Omega
(A)\right) ^{\vee m}\vee f^{\vee n}\right) $ for all $z\in \mathcal{H}$, we
obtain \newline
$\exp \left( \Omega (A)+f\right) =\exp \Omega (A)\vee \exp f$. For the
proofs given in this paper the definition $\Phi (A,f)=\exp \Omega (A)\vee
\exp f$ is sufficient. We have not used the identification with $\exp \left(
\Omega (A)+f\right) $.
\end{remark}

\subsection{Series expansion\label{series}}

In this Appendix we give a supplementary proof for the inner products (\ref
{h17}) and (\ref{h19}) using a series expansion.

For $H\in \mathcal{S}_{2}(\mathcal{H})$ we obtain from (\ref{h3a}) $\left\|
H^{\vee 2n}\right\| \leq \frac{(2n)!}{2^{n}}\left\| H\right\| ^{2n}$. Hence
the exponential series $\exp H=1+H+\frac{1}{2!}H^{\vee 2}+\frac{1}{3!}%
H^{\vee 3}+...$ converges within $\mathcal{S}(\mathcal{H})$ with the norm
estimate $\left\| \exp H\right\| ^{2}\leq \sum_{n=0}^{\infty }\left( \frac{1%
}{n!}\right) ^{2}\frac{(2n)!}{2^{n}}\left\| H\right\| ^{2n}<\infty $ if $%
\left\| H\right\| ^{2}<1/2.$ The mapping $H\in \mathcal{S}_{2}(\mathcal{H}%
)\rightarrow \exp H\in \mathcal{S}(\mathcal{H})$ is therefore an analytic
function within the open ball $\left\{ H\mid \left\| H\right\| <1/\sqrt{2}%
\right\} $. As a consequence of Lemma \ref{iso} we know that $\exp \Omega
(A)\in \mathcal{S}(\mathcal{H})$ if
\begin{equation*}
A\in \mathbf{B}_{1}:=\left\{ A\mid A\in \mathcal{L}_{2sym}(\mathcal{H}%
),\,\left\| A\right\| _{HS}<1\right\} \subset \mathbf{D}_{1}\subset \mathcal{%
L}_{2sym}(\mathcal{H})
\end{equation*}
and $A\in \mathbf{B}_{1}\rightarrow \exp \Omega (A)\in \mathcal{S}(\mathcal{H%
})$ is analytic. The ball $\mathbf{B}_{1}$ is an open subset of the convex
open set $\mathbf{D}_{1}\subset \mathcal{L}_{2sym}(\mathcal{H})$.

For the subsequent calculations we use a diagonalization of the symmetric
tensors $\Omega (A)$.

\begin{lemma}
\label{sym}Let $A\in \mathcal{L}_{2sym}(\mathcal{H})$ then $\Omega (A)$ has
a representation
\begin{equation}
\Omega (A)=\frac{1}{2}\sum_{\mu =0}^{\infty }\alpha _{\mu }\,f_{\mu }\vee
f_{\mu },  \label{b1}
\end{equation}
where $\left\{ f_{\mu }\right\} $ is a set of orthonormal vectors in $%
\mathcal{H}$ and the $\alpha _{\mu }$ are complex numbers such that the
series $\sum_{\mu }\left| \alpha _{\mu }\right| ^{2}=\left\| \Omega
(A)\right\| ^{2}=\frac{1}{2}\left\| A\right\| _{HS}^{2}$ converges. The
vectors $\left\{ f_{\mu }\right\} $ can be chosen such that $\alpha _{\mu
}\geq 0$.
\end{lemma}

\begin{proof}
See Lemma 2 of \cite{KMTP:1967}. The corresponding representation of the
operator $A$ is
\begin{equation}
Af=\sum_{\mu =0}^{\infty }\alpha _{\mu }f_{\mu }\left\langle f_{\mu }\mid
f\right\rangle =\sum_{\mu =0}^{\infty }\alpha _{\mu }f_{\mu }\left( f_{\mu
}^{\ast }\mid f\right) .  \label{b2}
\end{equation}
A direct proof of the representation (\ref{b2}) follows from Sect. 2.2 of
\cite{Gelfand4:1964}.
\end{proof}

In the main part of this subsection we calculate the function
\begin{equation}
\Phi (\bar{A},B,f):=\left( \exp \Omega (A)\mid \exp \Omega (B)\vee \exp
f\right) =\left\langle \exp \Omega (\bar{A})\mid \exp \Omega (B)\vee \exp
f\right\rangle  \label{b3}
\end{equation}
for $A,\,B\in \mathbf{B}_{1}$ and $f\in \mathcal{H}$. This function is
antianalytic in $A$ (analytic in $\bar{A}$) and analytic in $B$; it is
uniquely determined by its values on the diagonal $B=A$. The tensor $\Omega
(A)$ has the representation (\ref{b1}) with the additional constraint $%
\left\| A\right\| _{HS}=\sum_{\mu }\left| \alpha _{\mu }\right| ^{2}=\frac{1%
}{2}\left\| A\right\| _{HS}<1/2$. That yields the product representations $%
\exp \Omega (A)=\prod_{\mu }\exp \left( \frac{1}{2}\alpha _{\mu }e_{\mu
}\vee e_{\mu }\right) $ and
\begin{eqnarray}
\Phi (\bar{A},A,f) &=&\prod_{\mu }\varphi (\overline{\alpha _{\mu }},\alpha
_{\mu },\gamma _{\mu })\quad \text{with}  \label{b4} \\
\varphi (\overline{\alpha _{\mu }},\alpha _{\mu },\gamma _{\mu })
&=&\sum_{k,m,n=0}^{\infty }\frac{1}{k!m!(2n)!}\left( \left( \frac{\alpha
_{\mu }}{2}e_{\mu }\vee e_{\mu }\right) ^{k}\mid \left( \frac{\alpha _{\mu }%
}{2}e_{\mu }\vee e_{\mu }\right) ^{m}\vee \left( \gamma _{\mu }e_{\mu
}\right) ^{2n}\right)  \notag \\
\gamma _{\mu } &=&(e_{\mu }\mid f).  \notag
\end{eqnarray}
The inner product vanishes unless $k=m+n$. The remaining sum
\begin{equation*}
\varphi (\overline{\alpha },\alpha ,\gamma )=\sum_{m,n=0}^{\infty }\frac{%
2^{-2m-n}\overline{\alpha }^{m+n}\alpha ^{m}\gamma ^{2n}}{(m+n)!m!(2n)!}%
(2m+2n)!
\end{equation*}
can be evaluated using the identity
\begin{equation*}
\sum_{m=0}^{\infty }\frac{(2m+2n)!}{(m+n)!m!}z^{m}=\frac{(2n)!}{n!}%
\sum_{k}\left( \frac{2n+1}{2}\right) _{k}\frac{(4z)^{k}}{k!}=\frac{(2n)!}{n!}%
(1-4z)^{-\frac{2n+1}{2}}
\end{equation*}
such that
\begin{equation*}
\varphi (\overline{\alpha },\alpha ,\gamma )=\left( 1-\left| \alpha \right|
^{2}\right) ^{-\frac{1}{2}}\sum_{n}\frac{1}{n!}2^{-n}(1-\left| \alpha
\right| ^{2})^{-n}\overline{\alpha }^{n}\gamma ^{2n}=\left( 1-\left| \alpha
\right| ^{2}\right) ^{-\frac{1}{2}}\exp \left[ \frac{1}{2}(1-\left| \alpha
\right| ^{2})^{-1}\overline{\alpha }\gamma ^{2}\right] .
\end{equation*}
Then
\begin{equation*}
\Phi (\bar{A},A,f)=\det \left( I-\bar{A}A\right) ^{-\frac{1}{2}}\exp \frac{1%
}{2}\left\langle f\mid (I-\bar{A}A)^{-1}\bar{A}f\right\rangle
\end{equation*}
follows. The function (\ref{b3}) can therefore be written as
\begin{equation}
\left( \exp \Omega (A)\mid \exp \Omega (B)\vee \exp f\right) =\det \left(
I-A^{+}B\right) ^{-\frac{1}{2}}\exp \frac{1}{2}\left\langle f\mid
(I-A^{+}B)^{-1}A^{+}f\right\rangle .  \label{b5}
\end{equation}
This identity yields the inner product (\ref{h17})
\begin{equation}
\left( \exp \Omega (A)\mid \exp \Omega (B)\right) =\left( \det {}_{\mathcal{H%
}}\left( I-A^{+}B\right) \right) ^{-\frac{1}{2}}=\left( \det {}_{\mathcal{H}%
}\left( I-BA^{+}\right) \right) ^{-\frac{1}{2}},  \label{b6}
\end{equation}
so far derived for $A,\,B\in \mathbf{B}_{1}$. But the calculations of (\ref
{b4}) are well defined for $A\in \mathbf{D}_{1}$, and the right hand side of
(\ref{b6}) is antianalytic for $A\in \mathbf{D}_{1}$ and analytic for $B\in
\mathbf{D}_{1}$. Hence the mapping $A\in \mathbf{B}_{1}\rightarrow \exp
\Omega (A)\in \mathcal{S}(\mathcal{H})$ can be analytically continued to $%
A\in \mathbf{D}_{1}$, and the inner product of two of these vectors is given
by (\ref{h17}).

We can now apply the arguments of the Appendix \ref{norm} to define the
ultracoherent vector $\Phi (A,f)=\exp \Omega (A)\vee \exp f$ for $A\in
\mathbf{D}_{1}$ and $f\in \mathcal{H}$. Using (\ref{id3}) we verify the
identity
\begin{equation*}
\left( \exp (h+f)\mid \exp \Phi (B,g)\right) =\left( \exp h\mid \exp \Phi
(B,Bf^{\ast }+g\right) \mathrm{e}^{\frac{1}{2}\left\langle f^{\ast }\mid
Bf^{\ast }\right\rangle +\left\langle f^{\ast }\mid g\right\rangle }
\end{equation*}
for $B\in \mathbf{D}_{1}$ and $f,g,h\in \mathcal{H}$. Since $\mathcal{S}%
_{coh}(\mathcal{H})$ is dense in $\mathcal{S}^{(\alpha )}(\mathcal{H}%
),\,\alpha \in (0,1)$, this identity and Lemma \ref{Focknorms} imply
\begin{equation}
\left( H\vee \exp f\mid \exp \Phi (B,g)\right) =\left( H\mid \exp \Phi
(B,Bf^{\ast }+g\right) \mathrm{e}^{\frac{1}{2}\left\langle f^{\ast }\mid
Bf^{\ast }\right\rangle +\left\langle f^{\ast }\mid g\right\rangle }
\label{b7}
\end{equation}
for all $H\in \mathcal{S}^{(\alpha )}(\mathcal{H}),\,\alpha \in (0,1)$.
Given $A\in \mathbf{D}_{1}$ the vector $\exp \Omega (A)$ is an element of $%
\mathcal{S}^{(\alpha )}(\mathcal{H})$ for some $\alpha \in (0,1)$. Choosing $%
H=\exp \Omega (A)$ we obtain from (\ref{b5}) and (\ref{b7})
\begin{equation}
\begin{array}{c}
\left( \exp \left( \Omega (A)+f\right) \mid \exp \left( \Omega (B)+g\right)
\right) =\det (I-A^{+}B)^{-\frac{1}{2}} \\
\times \mathrm{e}^{\frac{1}{2}\left\langle f^{\ast }\mid Bf^{\ast
}\right\rangle +\left\langle f^{\ast }\mid g\right\rangle }\exp \frac{1}{2}%
\left\langle (Bf^{\ast }+g)\mid (I-A^{+}B)^{-\frac{1}{2}}A^{+}(Bf^{\ast
}+g)\right\rangle \\
=\exp \left( \frac{1}{2}\left\langle f^{\ast }\mid Cf^{\ast }\right\rangle
+\left\langle f^{\ast }\mid (I-BA)^{-1}g\right\rangle +\frac{1}{2}%
\left\langle g\mid Dg\right\rangle \right)
\end{array}
\label{b8}
\end{equation}
with the operators (\ref{h20}). Hence we have given another proof of (\ref
{h19}).


\begin{thebibliography}{99}
\bibitem{AY:1982}  H.~Araki and S.~Yamagami. \newblock On quasi-equivalence
of quasifree states of the canonical commutation relations.
\newblock {\em
Publ. RIMS, Kyoto Univ.}, 18(2):283--338, 1982.

\bibitem{BSZ:1992}  J.~C. Baez, I.~E. Segal, and Z.~Zhou.
\newblock {\em {Introduction to Algebraic and Constructive Quantum Field
  Theory}}. \newblock Princeton University Press, Princeton, 1992.

\bibitem{Banerjee/Kupsch:2005}  S.~Banerjee and J.~Kupsch. \newblock %
Applications of canonical transformations.
\newblock {\em J. Phys. A: Math.
Gen.}, 38:5237--5252, 2005.

\bibitem{Bargmann:1961}  V.~Bargmann.
\newblock {On a Hilbert space of analytic functions and an associated integral
  transform, Part I}. \newblock {\em Comm. Pure Appl. Math.}, 14:187--214,
1961.

\bibitem{Bargmann:1970}  V.~Bargmann.
\newblock {Group representations on
Hilbert spaces of analytic functions}. \newblock In R.~P. Gilbert and R.~G.
Newton, editors, \emph{Analytic Methods in Mathematical Physics}, pages
27--63, New York, 1970. Gordon and Breach.

\bibitem{Berezin:1966}  F.~A. Berezin.
\newblock {\em {The Method of Second
Quantization}}. \newblock Academic Press, New York, 1966.

\bibitem{Friedrichs:1953}  K.~O. Friedrichs.
\newblock {\em {Mathemathical
Aspects of the Quantum Theory of Fields}}. \newblock Interscience, New York,
1953.

\bibitem{Gelfand4:1964}  I.~M. Gelfand and N.~Ya. Vilenkin.
\newblock {\em {Generalized Functions. Vol. 4. Applications of Harmonic
  Analysis}}. \newblock Academic Press, New York, 1964.

\bibitem{Guichardet:1972}  A.~Guichardet.
\newblock {\em Symmetric {Hilbert}
spaces and related topics}. \newblock Lect. Notes in Math., Vol. 261.
Springer, Berlin, 1972.

\bibitem{Hall:2000}  B.~C. Hall. \newblock Holomorphic methods in analysis
and mathematical physics. \newblock {\em Contemporary Mathematics},
260:1--59, 2000. \newblock quant-ph/9912054 v2.

\bibitem{Hille/Phillips:1957}  E.~Hille and R.~S. Phillips.
\newblock {\em
{Functional Analysis and Semigroups}}. \newblock Amer. Math. Soc., 1957.

\bibitem{Ito/Hiroshima:2004}  K.~R. Ito and F.~Hiroshima. \newblock Local
exponents and infinitesimal generators of canonical transformations on boson
fock spaces.
\newblock {\em Infin. Dimens. Anal. Quantum Probab.
Relat. Top.}, 7:547--571, 2004. \newblock arXiv:math-ph/0309044.

\bibitem{Itzykson:1967}  C.~Itzykson.
\newblock {Remarks on boson
commutation rules}. \newblock {\em Commun. Math. Phys.}, 4:92--122, 1967.

\bibitem{Ken:1988}  T.~A.~B. Kennedy and D.~F. Walls. \newblock Squeezed
quantum fluctuations and macroscopic quantum coherence.
\newblock {\em Phys.
Rev. A}, 37:152--157, 1988.

\bibitem{Kim:1993}  M.~S. Kim and V.~Bu\u{z}ek. \newblock Photon statistics
of superposition states in phase-sensitive reservoirs.
\newblock {\em Phys.
Rev. A}, 47:610--619, 1993.

\bibitem{KMS:1975}  P.~Kramer, M.~Moshinsky, and T.~H. Seligman. \newblock %
Complex extensions of canonical transformations and quantum mechanics. %
\newblock In E.~M. Loebl, editor, \emph{Group Theory and Its Applications.
Vol. {III}}, pages 250--332. Academic Press, New York, 1975.

\bibitem{KMTP:1967}  P.~Kristensen, L.~Mejlbo, and E.~Thue Poulsen.
\newblock {Tempered distributions in infinitely many dimensions. III. Linear
  transformations of field operators}. \newblock {\em Commun. Math. Phys.},
6:29--48, 1967.

\bibitem{Kupsch/Smolyanov:2000a}  J.~Kupsch and O.~G. Smolyanov.
\newblock
{Bogolyubov transformations in Wiener-Segal-Fock space}.
\newblock {\em
Math. Notes}, 68(3/4):409--414, 2000.

\bibitem{Kupsch/Smolyanov:2000b}  J.~Kupsch and O.~G. Smolyanov.
\newblock {Realizations of unitary transformations generating the Bogoliubov
  transformations in spaces of the Wiener-Segal-Fock type}.
\newblock {\em
Dokl. Math.}, 61:169--173, 2000.

\bibitem{Kupsch/Smolyanov:2000}  J.~Kupsch and O.~G. Smolyanov. \newblock %
Hilbert norms for graded algebras. \newblock {\em Proc. Amer. Math. Soc.},
128:1647--1653, 2000. \newblock funct-an/9712005.

\bibitem{Nielsen:1991}  T.~T. Nielsen.
\newblock {\em {Bose Algebras: The
Complex and Real Wave Representations}}. \newblock Lect. Notes in Math.
1472. Springer, Berlin, 1991.

\bibitem{Ottesen:1995}  J.~T. Ottesen.
\newblock {\em {Infinite Dimensional
Groups and Algebras in Quantum Physics}}. \newblock Springer, Berlin, 1995. %
\newblock {Lect. Notes Phys. Vol. m 27}.

\bibitem{Parthasarathy:1992}  K.~R. Parthasarathy.
\newblock {\em
{Introduction to Quantum Stochastic Calculus}}. \newblock Birkh{\"a}user,
Basel, 1992.

\bibitem{Ruijsenaars:1978}  S.~N.~M. Ruijsenaars.
\newblock {On Bogoliubov
transformations. II. The general case}. \newblock {\em Ann. Phys. (N.Y.)},
116:105--134, 1978.

\bibitem{Segal:1962}  I.~E. Segal. \newblock Mathematical characterization
of the physical vacuum for a linear {Bose-Einstein} field.
\newblock {\em
Illinois J. Math.}, 6:500--523, 1962.

\bibitem{Shale:1962}  D.~Shale. \newblock Linear symmetries of free boson
fields. \newblock {\em Trans. Amer. Math. Soc.}, 103:149--167, 1962.

\bibitem{Siegel:1943}  C.~L. Siegel. \newblock Symplectic geometry. %
\newblock {\em Amer. J. Math.}, 65:1--86, 1943.

\bibitem{SSM:1988}  R.~Simom, E.~C.~G. Sudarshan, and N.~Mukunda. \newblock %
Gaussian pure states in quantum mechanics and the symplectic group. %
\newblock {\em Phys. Rev. A}, 37:3028--3038, 1988.

\bibitem{Slowikowski:1988}  W.~Slowikowski.
\newblock {Ultracoherence in
Bose algebras}. \newblock {\em Adv. Appl. Math.}, 9:377--427, 1988.
\end{thebibliography}

\end{document}